\documentstyle[11pt,newpasp,twoside,psfig]{article}
\def\edcomment#1{\iffalse\marginpar{\raggedright\sl#1\/}\else\relax\fi}
\reversemarginpar

\newcommand{\HI}{H\thinspace\protect\footnotesize
I\protect\normalsize}

\newcommand{\B}{{$B$}}

\newcommand{\II}{{$I_c$}}
\newcommand{\J}{{$J$}}
\newcommand{\HH}{{$H$}}
\newcommand{\K}{{$K_s$}}
\newcommand{\tfr}{Tully\,--\,Fisher relation}

\newcommand{\kms}{\,km\,s$^{-1}$}
\newcommand{\etal}{{et~al.\, }}
\newcommand{\cf}{{cf.\, }}
\newcommand{\eg}{{e.g.\, }}         
\newcommand{\ie}{{i.e.\, }}         

\def\bfs{{\bf s}}
\def\bfS{{\rm S}}

\def\deg{\hbox{$^\circ$}}
\def\arcmin{\hbox{$^\prime$}}
\def\arcsec{\hbox{$^{\prime\prime}$}}
\def\micron{\hbox{$\mu$m}}

\begin{document} 
\title{The Universe behind the Milky Way}
 
\author{Ren\'ee C. Kraan-Korteweg\altaffilmark{1} and 
Ofer Lahav\altaffilmark{2}}

\affil{$^{1}$Depto. de Astronom\1a, Universidad de Guanajuato, Apartado
Postal 144, Guanajuato GTO 36000, Mexico\\
$^2$
Institute of Astronomy, Madingley Road, Cambridge CB3 OHA, UK
}

 
\begin{abstract}
Due to the foreground extinction of the Milky Way, galaxies appear
increasingly fainter the closer they lie to the Galactic Equator,
creating a "zone of avoidance" of about 25\% in the distribution of
optically visible galaxies. A "whole-sky" map of galaxies is
essential, however, for understanding the dynamics in our local
Universe, in particular the peculiar velocity of the Local Group with
respect to the Cosmic Microwave Background and velocity flow fields
such as in the Great Attractor region.

Various dynamically important structures behind the Milky Way have
only recently been made ``visible'' through dedicated deep surveys at
various wavelengths. The wide range of observational searches
(optical, near infrared, far infrared, radio and X-ray) for galaxies
in the Zone of Avoidance are reviewed, including a discussion on the
limitations and selection effects of these partly complementary
approaches. The uncovered and suspected large-scale structures are
summarized. Reconstruction methods of the density field in the Zone of
Avoidance are described and the resulting predictions compared with
observational evidence.  The comparison between reconstructed density
fields and the observed galaxy distribution allow derivations of the
density and biasing parameters $\Omega_0$ and $b$.
\end{abstract}


\section{Introduction}
The unveiling of the galaxy distribution behind the Milky Way has
turned into a research field of its own in the last ten years (\eg
``Unveiling Large-Scale Structures behind the Milky Way'', ASP
Conf. Ser. 67, eds.  Balkowski \& Kraan-Korteweg, 1994, and ``Mapping
the Hidden Universe'', ASP Conf. Ser., eds. Kraan-Korteweg \etal 2000,
in press).  Why is it of interest to know the galaxy distribution
behind the Milky Way, and why is it not sufficient to study galaxies
and their large-scale distribution away from the foreground
``pollution'' of the Milky Way?  To understand the dynamics in the
nearby Universe and answer the question whether the the dipole in the
Cosmic Microwave Background (CMB) and other velocity flow fields (\eg
towards the Great Attractor) can be fully explained by the clumpy
galaxy/mass distribution, whole-sky coverage is essential. The lack of
data in large areas of the sky -- where the size of the gap due to the
Milky Way depends on the wavelength at which galaxies are sampled --
constitutes a severe restriction in solving these questions.

Based on various dedicated observational programs, using nearly all
the bands of the electromagnetic spectrum, and the charting of
large-scale structures and flow fields in the Zone of Avoidance (ZOA)
from statistical reconstructions, a lot of progress has been
achieved. In this review, we will give a status report on all of the
above approaches. After a historic perspective on the ZOA
(Sect.~\ref{zoa}), the cosmological questions for which the unveiling
of the ZOA are most relevant are described (Sect.~\ref{const}). In
Sect.~\ref{ext}, a discussion on the current knowledge of the
foreground extinction is presented. This is followed by a description
of the various observational multi-wavelength techniques that are
currently being employed to uncover the galaxy distribution in the ZOA
such as deep optical searches (Sect.~\ref{opt}), near-infrared and
far-infrared surveys (Sect.~\ref{nir}, and \ref{fir}),
systematic blind radio surveys (Sect.~\ref{hi}) and searches for
hidden massive X-ray clusters (Sect.~\ref{Xray}). For each method, the
different limitations and selection effects and results are
presented. The various statistical reconstruction methods are reviewed
and results of the density field in the ZOA compared with
observational data (Sect.~\ref{recon}).

\section{The Zone of Avoidance}\label{zoa}
A first reference to the Zone of Avoidance (ZOA), or the ``Zone of few
Nebulae'' was made in 1878 by Proctor, based on the distribution of
nebulae in the ``General Catalogue of Nebulae'' by Sir John Herschel
(1864). This zone becomes considerably more prominent in the
distribution of nebulae presented by Charlier (1922) using data from
the ``New General Catalogue'' by Dreyer (1888, 1895). These data also reveal
first indications of large-scale structure: the nebulae display a very
clumpy distribution. Currently well-known galaxy clusters such as 
Virgo, Fornax, Perseus, Pisces and Coma are easily recognizable even
though Dreyer's catalog contains both Galactic and extragalactic objects
as it was not known then that the majority of the nebulae actually
are external stellar systems similar to the Milky Way. Even more
obvious in this distribution, though, is the absence of galaxies around
the Galactic Equator. As extinction was poorly known at that time, no
connection was made between the Milky Way and the ``Zone of few
Nebulae''.

A first definition of the ZOA was proposed by Shapley (1961) as the
region delimited by ``the isopleth of five galaxies per square degree
from the Lick and Harvard surveys'' (compared to a mean of 54
gal./sq.deg. found in unobscured regions by Shane \& Wirtanen,
1967). This ``Zone of Avoidance'' used to be ``avoided'' by
astronomers interested in the extragalactic sky because of the lack of
data in that area of the sky and the inherent difficulties in
analyzing the few obscured galaxies known there.

Merging data from more recent galaxy catalogs, \ie the Uppsala General
Catalog UGC (Nilson 1973) for the north ($\delta \ge -2\fdg5$), the
ESO Uppsala Catalog (Lauberts 1982) for the south ($\delta \le
-17\fdg5$), and the Morphological Catalog of Galaxies MCG
(Vorontsov-Velyaminov \& Archipova 1963-74) for the strip inbetween
($-17\fdg5 < \delta < -2\fdg5$), a whole-sky galaxy catalog can be
defined. To homogenize the data determined by different groups from
different survey material, the following adjustments have to be
applied to the diameters: ${D = 1.15 \cdot D_{\rm UGC}, D = 0.96 \cdot
D_{\rm ESO}}$ and ${D = 1.29 \cdot D_{\rm MCG}}$ (see Fouqu\'e \&
Paturel 1985, Lahav 1987). According to Hudson \& Lynden-Bell (1991)
this ``whole-sky'' catalog then is complete for galaxies larger than
${D} =1\farcm3$.

\begin{figure}[ht]
\begin{center}
\hfil \psfig{file=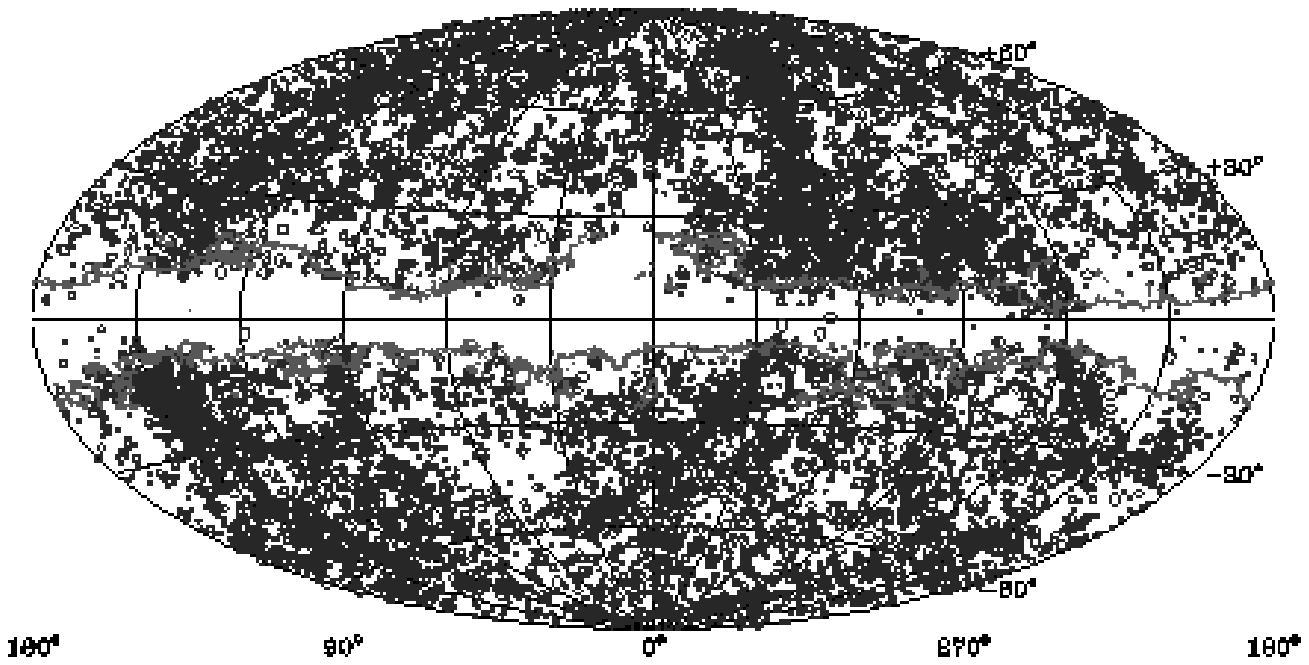,width=12cm} \hfil
\caption
{Aitoff equal-area projection in Galactic coordinates of galaxies with
${D}\ge1\farcm$3. The galaxies are diameter-coded: small circles
represent galaxies with $1{\farcm}3 \le {D} < 2\arcmin$, larger
circles $2\arcmin \le {D} < 3\arcmin$, and big circles ${D}
\ge 3\arcmin$. The contour marks absorption in the blue of ${A_B}
= 1\fm0$ as determined from the Schlegel \etal (1998) dust extinction
maps.  The displayed contour surrounds the area where the galaxy
distribution becomes incomplete (the ZOA) remarkably well.}
\label{ait}\protect
\end{center}
\end{figure}

The distribution of these galaxies is displayed in Galactic
coordinates in Fig.~\ref{ait} in an equal-area Aitoff projection
centered on the Galactic Bulge ($\ell = 0\deg, b = 0\deg$). The
galaxies are diameter-coded, so that structures relevant for the
dynamics in the local Universe stand out accordingly.  Figure~\ref{ait}
clearly displays the irregularity in the distribution of galaxies in
the nearby Universe such as the Local Supercluster visible as a great
circle (the Supergalactic Plane) centered on the Virgo cluster at
$\ell=284\deg, b=74\deg$, the Perseus-Pisces chain (PP) bending into the
ZOA at $\ell=95\deg$ and $\ell=165\deg$, the general overdensity in
the Cosmic Microwave Background dipole direction
($\ell=276\deg,b=30\deg$; Kogut \etal 1993) and the general galaxy
overdensity in the Great Attractor region (GA) centered on $\ell=320\deg,
b=0\deg$ (Kolatt \etal 1995) with the Hydra ($270\deg,27\deg$), Antlia
($273\deg,19\deg$), Centaurus ($302\deg,22\deg$) and Pavo
($332\deg,-24\deg$) clusters.

Most conspicuous in this distribution is, however, the very broad,
nearly empty band of about 20$\deg$ width. As optical galaxy catalogs are
limited to the largest galaxies they become increasingly incomplete
close to the Galactic Equator where the dust thickens. This diminishes
the light emission of the galaxies and reduces their visible
extent. Such obscured galaxies are not included in diameter- or
magnitude-limited catalogs because they appear small and faint -- even
though they might be intrinsically large and bright. A further
complication is the growing number of foreground stars close to the
Galactic Plane (GP) which fully or partially block the view of galaxy
images. 

Comparing this ``band of few galaxies'' with the currently available
100\micron\ dust extinction maps of the DIRBE experiment (Schlegel
\etal 1998; see Sect.~\ref{ext}), we can see that the ZOA -- the area where
the galaxy counts become severely incomplete -- is described almost
perfectly by the absorption contour in the blue ${A_B}$ of $1\fm0$
(where ${A_B} = 4.14 \cdot E(B-V)$; Cardelli \etal 1989). This contour
matches the by Shapley (1961) defined ZOA closely.
 
\section{The ZOA as an obstacle for cosmological studies}\label{const}

\subsection{Large-scale structures} \label{LSS}

Enormous effort and observation time has been devoted in the last
decades to establish the extragalactic large-scale structure. 
From the various redshift slices, 3-dimensional pictures evolve that
distribute galaxies predominantly in clusters, sheets and filaments,
leaving large areas devoid of luminous matter. These filaments,
respectively their sizes, carry information on the conditions and
formation processes of the early Universe, providing important
constraints which must be reproduced in cosmological models. 

Many of the known nearby large-scale structures are, however, bisected
by the Galactic Plane - such as the Local Supercluster, the
Perseus-Pisces chain, and the Great Attractor (see
Fig.~\ref{ait}). What is their true extent and their mass? It is
curious that the two major superclusters in the local Universe, \ie
Perseus-Pisces and the Great Attractor overdensity, lie at similar
distances on opposite sides of the Local Group (LG) -- both partially
obscured by the ZOA. Which one of the two is dominant in the
tug-of-war on the LG? Do these features continue across the Galactic
Plane and are there other massive structures hidden in the ZOA for
which so far no indication exists?

What is the size of the largest coherent structures? Are, for
instance, the Great Wall and the Perseus-Pisces chain connected across
the ZOA as suggested in 1982 by Giovanelli \& Haynes (but see Marzke
\etal 1996), indicating structures of over $200~h_{50}^{-1}$~Mpc. The
latter would be incompatible with the angular extent over which
fluctuations -- the seeds of current large-scale structures -- have
been measured in the CMB. To answer these questions, superclusters
need to be fully mapped across the ZOA.

\subsection{Dipole determinations}\label{dipol}
Filling the ZOA also is paramount with respect to the evaluation of
the peculiar velocity of the Local Group. The dipole anisotropy in the
CMB radiation is explained by a peculiar motion of the LG relative to
the CMB of ${\vec v_p} = 627$~\kms\ towards $\ell = 276\deg, b =
30\deg$ (Kogut \etal 1993). This motion arises from the net
gravitational attraction on the LG due to the irregular distribution
of matter in the Universe.  Reproducing the vector measured in the CMB
radiation with the LG motion determined from the matter distribution
(direction as well as convergence distance) will lead to constraints
on the cosmological parameter $\Omega_0$.

The determination of the gravity field at the position of the LG, \ie
velocity and direction of the peculiar motion, requires whole-sky
coverage. Kolatt \etal (1995) have shown, for instance, that the mass
distribution within $\pm20\deg$ of the ZOA -- as derived
from theoretical reconstructions of the density field (see
Sect.~\ref{recon}) -- is crucial to the derivation of the
gravitational acceleration of the LG: the direction of the motion
measured within a volume of 6000~\kms\ will change by $31\deg$ when
the (reconstructed) mass within the ZOA is included.  The results
derived so far for the apex of the LG motion, as well as the distance
at which convergence is attained, still are controversial. The lack of
data for the ZOA remains one of the main uncertainty in current dipole
determinations (\eg Rowan-Robinson \etal 2000).

Most dipole determinations have assumed a uniformly filled ZOA (\eg by
Poissonian statistics) or have used cloning methods which transplant
the fairly well-mapped regions adjacent to the ZOA into the ZOA, or a
spherical harmonic analysis (Sect.~\ref{recon}). All procedures are
unsatisfactory, because inhomogenous data coverage, incorrect
assumptions on the galaxy distribution in the ZOA, and/or false
assumptions on the ZOA mask to be filled will introduce nonexisting
flow fields.  Care should therefore be taken on how to extrapolate the
galaxy density field across the ZOA. Obviously, a reliable consensus
on the galaxy distribution in the ZOA is important to minimize these
uncertainties.

In this context, not only the identification of unknown and suspected
clusters, filaments and voids are relevant, but also the detection of
nearby smaller entities. In linear theory, the peculiar velocity of
the LG is propertional to the net gravity field which can be
determined from the sum of the masses of all galaxies divided by the 
distance squared:
$${\vec v}_p \propto {\Omega_0^{0.6} \over b} \sum
{{\cal M}_{i} \over r_{i}^2} \, {\hat {\bf r_i}},$$ 
where $\Omega_0$ is the density parameter and $b$ the bias parameter.
Since gravity as well as the flux of a galaxy decrease with $r^{-2}$, the
direction and amplitude of the peculiar velocity can be determined
directly from, for instance, the sum of the {\sl apparent magnitudes}
of the galaxies in the sky under the assumption of constant
mass-to-light ratio
$${\vec v}_p \propto \sum_{\rm i} 10^{-0.4{\rm m}} \, {\hat {\bf
r_i}}.$$ 
This has important implications and suggests, for instance, that the
galaxy CenA with an absorption-corrected magnitude of $B^o = 6\fm1$
exerts a stronger luminosity-indicated gravitational attraction on the
Local Group than the whole Virgo cluster. The problem whether galaxies
trace the mass is inherent to all cumulative dipole
determinations. These calculations also predict that the 8 apparently
brightest galaxies -- which are all nearby ($v < 300$~\kms) -- are
responsible for 20\% of the total dipole as determined from optically
known galaxies within $v \la 6000$~\kms. Hence, a major part of the
peculiar motion of the LG is generated by a few average, but nearby
galaxies. Note, however, that for nearby objects non-linear dynamics
has to be taken into account.

In this sense, the detection of nearby galaxies or galaxy groups
hidden by the obscuration layer of the Galaxy can be as important as
the detection of entire clusters at larger distances. The expectation
of finding additional nearby galaxies in the ZOA is not
unrealistic. Six of the nine apparently brightest galaxies
(extinction-corrected) are located in the ZOA: IC342, Maffei 1, Maffei
2, NGC4945, CenA and the recently discovered galaxy Dwingeloo 1
(Kraan-Korteweg \etal 1994b). In the absence of Galactic extinction
both Maffei 1 and IC 342 would subtend angles as large as the full
Moon (McCall \& Buta 1996).

\subsection{Dynamics of the Local Group} \label{dyn}
It is commonly believed that the Local Group (LG) of galaxies is
dominated by the Milky Way and Andromeda (M31).  A discovery of an
unknown Andromeda-like galaxy behind the Milky Way may therefore
dramatically change our understanding of the LG dynamics.

Kahn \& Woltjer (1959) and Lynden-Bell (1982) have shown that the
distances and motions of nearby galaxies may be used to constrain the
total mass of the pair Milky Way and M31 and the dynamical age of the
Local Group.  The method, known as ``Local Group timing'', is based upon
the assumption that M31 and the Milky Way separated from the Hubble
flow soon after formation and behaved dynamically like an
isolated binary ever since, with the consequence that the Milky Way
has turned in its orbit and is now falling back toward M31.  However,
it seems that the dominant members of the IC 342/Maffei Group may be
massive enough and near enough to the Local Group to have had an
influence on its dynamical history (McCall 1989; Valtonen et
al. 1993), thereby calling into question the binary hypothesis of LG
timing.  Peebles (1990) and Dunn \& Laflamme (1995) extended this
approach by tracing the orbits of the LG galaxies back in time, under
the ``least action principle''. These analyses rely heavily on having
a full census of LG galaxies and on accurate distances to them.
Therefore, better mapping of the ZOA is important in two ways: (a)
discovering new members of the LG and neighbouring groups (or ruling
out their existence); (b) measuring the intrinsic properties of
galaxies, properly corrected for extinction, to accurately 
estimate their distances.

\subsection{Cosmic flow fields} \label{flow}
Density enhancements locally decelerate the uniform expansion field,
as observed within our own Local Supercluster, resulting in systematic
streaming motions over and above the uniform expansion field. These
flows, on the other hand, can be exploited to map the mass density
field independent of the galaxy distribution and/or an assumption on
the mass-to-light ratio using peculiar velocities of galaxies, 
${\vec v_p} = {\vec v_{\rm obs}} - {\vec v_{\rm H}}$, for which distance
determinations independent of redshift are available. The latter
can be obtained via the Tully-Fisher relation for spiral galaxies
(Tully and Fisher 1977) or the $D_n - \sigma$ relation for elliptical
galaxies (Lynden-Bell \etal 1988). Note though, that only the radial
component of the peculiar motion of a galaxy can be measured. The
reconstruction of such potential fields and density fields have the
advantage that they can locate hidden mass overdensities
even if ``unseen''. These methods are therefore of particular interest
for ZOA research, as these potential fields provide information on
the mass distribution behind the Milky Way without having access to
the real data on the galaxies hidden there.

Based on these considerations, Dressler \etal (1987) interpreted a
systematic infall pattern from the peculiar velocities of about 400
elliptical galaxies as being due to a hypothetical Great Attractor
with a mass of $\sim 5\times 10^{16}{\cal M}_\odot$ at a position in
redshift space of $(\ell,b,v) = (307\deg,9\deg,\sim4500$~\kms)
(Lynden-Bell \etal 1988).  A more recent study by Kolatt \etal (1995),
based on a larger data set (elliptical {\sl and} spiral galaxies) and
the potential reconstruction method POTENT (see Sect.~\ref{recon} and
Fig.~\ref{kolatt}) place the center of the GA right behind the Milky
Way. Recent consensus is that the GA is an extended region ($\sim
40\deg \times 40\deg$) of moderately enhanced galaxy density. Although
there is a considerable excess of optical galaxies and IRAS-selected
galaxies in this region (see Fig.~\ref{ait} and Fig.~\ref{BTP}), no
dominant cluster or central peak can been seen. But the central part
of the GA is hidden by the Milky Way.

Large surveys (\eg Mathewson \etal 1992 for field galaxies, Mould
\etal 1991, Han 1992 for clusters) have resulted in a large
collection of peculiar velocities of galaxies, put together in the
Mark III catalog of peculiar velocities (Willick \etal 1997).  A
large effort to obtain peculiar motions of a sample of uniform sky
coverage (excluding, however, the ZOA at $|b| \ge 10\deg$) is presented by
Giovanelli \etal (1997) and da Costa \etal (1996).  This has opened up
a new field in cosmology, namely the dynamics of cosmic flows or
large-scale dynamics.

\section{Extinction by the Milky Way}\label{ext}

A crucial step in exploring the extragalactic sky behind the ZOA is a
detailed understanding of our own Galaxy, since it would otherwise remain
impossible to disentangle newly unveiled clustering from the
patchiness of the foreground extinction and, moreover, to correct the
observed parameters of detected galaxies for the absorption effects.
This requires (a) a high-resolution, well-calibrated map of the
foreground extinction and (b) a clear understanding how Galactic
extinction affects the observed galaxy parameters. Cameron (1990)
investigated the latter in the optical by artificially obscuring
high-latitude galaxies, an approach which really needs to be refined
and to be explored at other wavelengths.

The Galactic foreground extinction is a function of wavelength.
According to Cardelli \etal (1989), the mean extinction at a given 
wavelength, ${A_\lambda}$, compared to the visual extinction, $A_V$, is:
$${{{A_\lambda}}/{{A_V}}} = a(1/\lambda) + {{b(1/\lambda)}/{R_V}}$$
(see Mathis 1990, for a
comprehensive overview on the interstellar dust in the Galaxy). The
interstellar extinction is a function of the ratio of total to
selective extinction, \eg $R_V$ [$\equiv {A_V/E(B-V)}$] and the 
actual value depends on the environment along the
line of sight through the Galaxy. For the diffuse interstellar medium
a standard value of $R_V=3.1$ applies, whereas a higher value of
$R_V$ is evident for lines of sight through dense molecular clouds
($R \ge 4$). But this ratio of total to selective extinction
generally is founded upon observations of stars in the Milky Way,
ignoring shifts in effective wavelengths which are known to depend
upon reddening and intrinsic color. For galaxies heavily reddened by
dust in the Milky Way, this approach can lead to significant errors in
both magnitudes and distances (McCall \& Armour 2000).  So the values
for $R_V$ are by no means generally applicable.

For years, the extinction maps by Burstein \& Heiles (1982) were the
standard. But they do not cover the ZOA ($|b|>10\deg$). These maps 
can, however, be extrapolated towards lower latitudes,
following the precepts of Burstein \& Heiles (1978, 1982)
using the Galactic \HI\ column densities $N_{\rm HI}$ (\eg from
Kerr \etal 1986 for the south, and Hartmann \&
Burton 1997 for the north) assuming a constant gas-to-dust ratio:
$$E(B-V) = \left({{N_{\rm HI}}\over{2.23 \cdot 10^{18}}}\right) \times 4.43 \cdot 10^{-4} - 0.055.$$
However, the gas-to-dust ratio does vary. Burstein \etal (1987)
report, for instance, an increase in the gas-to-dust ratio for a
region in the southern Milky Way ($230\deg \le \ell \le 310\deg$ and
$-20\deg \le b \le 20\deg$) of up to a factor of 2, implying severe
overstimates of the extinction.  Moreover, close to the Galactic Plane
($|b| \la 2\deg$), the Galactic \HI\ line might be saturated, leading
to an underestimate of the true extinction. At these latitudes
though, the Galactic CO (Dame \etal 1987) can be used as a tracer of
extinction.

The recently published 100\micron\ extinction maps from the DIRBE
experiment (Schlegel \etal 1998) give an improved estimate of the
foreground extinction because they provide a direct measure of the
dust column density, and because these maps have better angular
resolution (6$\farcm$1 compared to $\sim 20-30\arcmin$ of the \HI\
maps). According to Schlegel et al., they are a factor 2 better at low
and moderate extinction compared to the Burstein \& Heiles
maps. However, as stated by Schlegel \etal themselves, the accuracy of
the DIRBE maps still needs to be established at low Galactic latitudes
($|b| \le 10\deg$).  Woudt (1998) found -- from photometry and
measurements of the Mg$_2$-index of 18 early type galaxies in the ZOA
-- that the extinction for moderate to high DIRBE reddenings is
systematically underestimated (by a factor of f=0.86).  As his new
calibration so far is based only on few galaxies in a small region of
the ZOA, it seems too early to incorporate these adjustments to the
DIRBE maps.  However, this study clearly illustrates the need for a
careful calibration of the DIRBE maps in the ZOA.

Although the knowledge on the Galactic foreground extinction has
improved enormously in the last 20 years, many open points still need
to be resolved before we can properly analyze the galaxy distribution
uncovered in the ZOA.  On the other hand, a lot can be learned about
total extinction from ZOA research itself.  As mentioned above, Woudt
(1998) has used photometry and spectroscopy of early-type galaxies in
the ZOA to obtain a first calibration of the DIRBE maps in the ZOA. He
now has started a program to pursue this systematically for the
southern Milky Way using a sample of about 300 early-type galaxies
distributed within the southern ZOA. Furthermore, Saito \etal (2000)
have used $B-I$ colors of over a hundred galaxy candidates at
extremely low latitudes ($|b|<1\deg$) in combination with CO 
emission to determine extinction estimates, and Temporin
\etal (2000) determined total extinctions from $BVRI$ photometry
towards galaxies in the region $29\deg < \ell < -14\deg, |b| <
11\deg$.  Similarly, the colors of galaxies identified in the ZOA with
the near infrared surveys DENIS and 2MASS (see Sect.~\ref{nir}) will
provide a huge data base to calibrate the DIRBE maps at low
Galactic latitudes.

\section{Optical surveys} \label{opt}

Systematic optical galaxy catalogs are generally limited to the
largest galaxies (typically with diameters D $\ga 1\arcmin$, \eg
Lauberts 1982). These catalogs become, however, increasingly
incomplete as the dust thickens, creating a ``Zone of Avoidance'' in
the distribution of galaxies of roughly 25\% of the
sky. Systematically deeper searches for partially obscured galaxies --
down to fainter magnitudes and smaller dimensions compared to existing
catalogs -- have been performed on existing sky surveys with the
aim of reducing this ZOA. Meanwhile, through the efforts of various
collaborations, nearly the whole ZOA has been surveyed and over 50000
previously unknown galaxies were discovered in this way. These
surveys are not biased with respect to any particular morphological
type and were able to identify important new large-scale structures in
and across the Milky Way.

\subsection{Early searches and results} 

One of the first attempts to detect galaxies in the ZOA was carried
out by B\"ohm-Vitense in 1956. She did follow-up observations in
selected fields in the GP in which Shane and Wirtanen (1954) found
objects that "looked like extragalactic nebulae" but were not believed
to be galaxies because they were so close to the dust equator. She
confirmed many galaxies and concluded that the obscuring matter in the
GP must be extremely thin and full of holes between
$\ell = 125\deg$ and $130\deg$.

Because extinction was known to be low in Puppis, Fitzgerald (1974)
performed a galaxy search in one field there ($\ell \sim 245\deg$)
and discovered 18 small and faint galaxies. Two years later Dodd \&
Brand (1976) examined 3 fields adjacent to this area ($\ell \sim
243\deg$) and detected another 29 galaxies.  Kraan-Korteweg \&
Huchtmeier (1992) observed these galaxies in \HI\ at Effelsberg and
identified an unknown nearby cluster at ($\ell,b,v) =
(245\deg,0\deg,\sim1500$~\kms). Including IRAS data in the analysis of
this cluster, it could be shown that its density is comparable to the
Virgo cluster and that this Puppis cluster may contribute a 
significant component to the motion of the LG (Lahav \etal 1993).

During a search for infrared objects, Weinberger \etal (1976) detected
two galaxy candidates near the Galactic Plane ($\ell \sim 88\deg$)
which Huchra \etal confirmed in 1977 to be the brightest members of a
galaxy cluster at 4200~\kms. This discovery led Weinberger (1980) to
start the first {\sl systematic} galaxy search. Using the POSS E
prints, he covered the whole northern GP ($\ell = 33\deg - 213\deg$)
in a thin strip $(|b| \le 2\deg)$. He found 207 galaxies, the
distribution of which is highly irregular: large areas disclose no
galaxies and the "hole" pointed out by B\"ohm-Vitense was verified, but
most conspicuous was a huge excess of galaxies around $\ell = 160\deg -
165\deg$. In 1984, Focardi \etal made the connection with large-scale
structures: they interpreted the excess as the possible continuation
of the Perseus-Pisces cluster across the GP to the cluster
A569.  Radio-redshift measurements by Hauschildt (1987) established
that the PP cluster at a mean redshift of $v = 5500$~\kms\ extends to
the cluster 3C129 in the GP ($\ell = 160\deg, b = 0\fdg1$). Additional
\HI\ and optical redshift measurements of Zwicky galaxies by Chamaraux
\etal (1990) indicate that this chain can be followed even further to
the A569 cloud at $v \sim 6000$~\kms\ on the other side of the ZOA.

These early searches proved that large-scale structure can be traced
to much lower Galactic latitudes despite the foreground obscuration
and its patchy nature which suggests clustering in the
galaxy distribution independent of large-scale structure. The above
investigations did confirm suspected large-scale features across the
GP through searches in selected regions and follow-up redshift
observations. To study large-scale structure systematically
broader latitude strips covering the whole Milky Way, respectively
the whole ZOA (see Fig.~\ref{ait}) are required.

\subsection{Status of systematic optical searches}\label{optsear}

Using existing sky surveys such as the first and second generation
Palomar Observatory Sky Surveys POSS I and POSS II in the north, and
the ESO/SRC (United Kingdom Science Research Council) Southern Sky
Atlas, various groups have performed systematic deep searches for
``partially obscured'' galaxies, \ie they catalogued galaxies down to
fainter magnitudes and smaller dimensions (${D} \ga 0\farcm1$) than
existing catalogs. Here, examination by eye remains the best
technique. A separation of galaxy and star images can not be
done as yet on a viable basis below $|b| \la 10\deg-15\deg$ by automated
measuring machines such as \eg COSMOS (Drinkwater \etal 1996) or APM
(Lewis \& Irwin 1996) and sophisticated extraction algorithms, nor
with the application of Artificial Neural Networks [ANN]. The latter
was tested by Naim (1995) who used ANN to identify galaxies with
diameters above 25$\arcsec$ at low Galactic latitudes ($b\sim
5\deg$). Galaxies could be identified using this algorithm, and
although an acceptable hit rate for galaxies of 80 -- 96\% could be
attained when ANN was trained on high latitude fields, the false
alarms were of equal order. Using low latitude fields as training
examples, the false alarms could be reduced to nearly zero but then
the hit rate was low ($\sim$ 30 - 40\%). The first attempts of using
ANN in the ZOA are encouraging but clearly need further development. So,
although surveys by eye are both tiring and time consuming, and
maybe not as objective, they currently still provide the best
technique to identify partially obscured galaxies in crowded star
fields.

Meanwhile, nearly the whole ZOA has been visually surveyed for
galaxies. The various surveyed regions are displayed in
Fig.~\ref{cor}. Details and results on the uncovered galaxy
distributions and the respective references are described below:

\begin{figure}[ht]
\begin{center}
\hfil \psfig{file=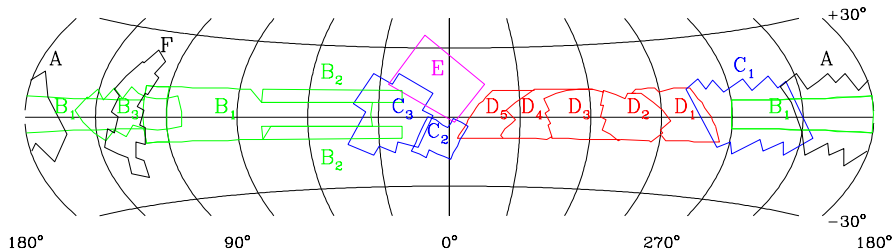,width=12cm} \hfil
\caption
{An overview of the different optical galaxy surveys in the ZOA centered
on the Galaxy. The labels identifying the search areas are explained 
in the text. Note that the surveyed regions cover the entire ZOA
defined by the foreground extinction level of ${A_B} = 1\fm0$ in
Fig.~1.}
\label{cor}
\end{center}
\end{figure}

{\bf Region A:} 
In order to trace the possible continuity across the Galactic Plane of
the southwestern spur of the PP complex Pantoja \etal (1994, 1997)
have searched for galaxy candidates on 29 POSS E prints with a 12 $x$
magnification. They identified 1480 galaxy candidates in the region delimited
by $4^h \le \alpha \le 8^h, 0\deg \le \delta \le 37\deg$, where the
declination range was optimized for \HI\ follow-up redshift
observations with the 305~m Arecibo radio telescope.

{\bf Regions B$_{1}$-B$_{3}$}: 
Using the first generation POSS I prints, and more recently also the
deeper POSS II films, Weinberger and collaborators in Austria have
expanded their optical galaxy searches in such a way that they now
cover nearly the whole northern Milky Way. They also searched the
Puppis region for comparative purposes. The information and data of
the B$_1$ region from POSS I plates can be found in Seeberger \etal
(1994), Seeberger \etal (1996), Lercher \etal (1996), Saurer \etal
(1997), Seeberger \etal (1998), and Marchiotto \etal 1999 for the
B$_2$ regions (also POSS I), and the B$_3$ region from POSS II fields
in Weinberger \etal 1999).  In total, they uncovered about 9500
galaxies in the northern ZOA. Their distribution shows a marked
overdensity at the suspected connection of the PP supercluster across
the Galactic Plane ($\l \sim 165\deg$). A comparison between the old
and new generation POSS fields found a dramatic increase in galaxy
numbers of about a factor of 3 for the deeper POSS II fields.

{\bf Regions C$_{1}$-C$_{3}$}:
Using the infrared film copies of the ESO/SRC survey Japanese groups
led by Saito investigated the ZOA in the longitude range $205 \la \ell
\la 260\deg, |b| < 10\deg$ (C$_1$: Saito \etal 1990, 1991). On 32
fields they charted over 7000 galaxies with $D \ge 0.1$~mm. As
anticipated, a large number of galaxies was detected in the low
opacity region of Puppis ($\ell \sim 245\deg$). Correlating the
density with the \HI-column density, they found indications of the
existence of various clusters.  They then continued studying regions
close to the Galactic bulge, \ie the Sagittarius/Galactic region
(C$_2$: $-7 \la \ell \la -16\deg, -19 \la b \la -1\deg$ by Roman \etal
1998), and the Aquila and Sagittarius region (C$_{3}$: $8 \la \ell
\la 47\deg, |b| \la 17\deg$ by Roman \etal 1996) uncovering a further
12500 galaxies in this very opaque region.  This data set is less
homogeneous as a variety of survey material had to be used to cover
this area.
 
\begin{figure}[ht]
\begin{center}
\hfil \psfig{file=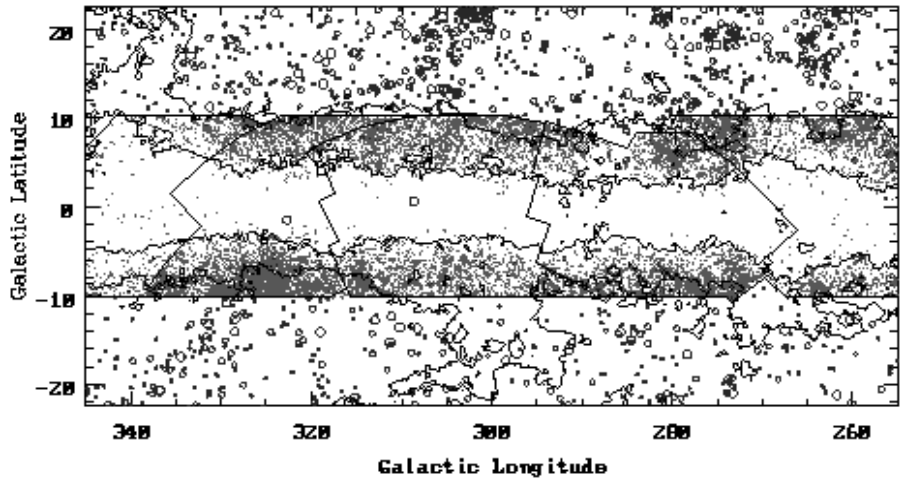,width=12cm} \hfil
\caption
{Distribution of Lauberts galaxies with $D \ge 1\farcm3$ (open circles
-- coded as in Fig.~1) and galaxies with $D \ge 12\arcsec$
(small dots) identified in the deep optical galaxy searches
D$_1$-D$_5$. The contours represent extinction levels of $A_{B} =
1\fm0$ and $3\fm0$. Note how the ZOA could be filled to $A_{B} =
3\fm0$ and that galaxy over- and underdensities uncorrelated with 
extinction can be recognized in this distribution.}
\label{D1_5}
\end{center}
\end{figure}

{\bf Regions D$_1$-D$_5$}: 
Since 1988, various groups led by Kraan-Korteweg have searched the
southern Milky Way between the Puppis region (C$_1$) and the Galactic
bulge region (C$_2$ \& E) using the IIIaJ film copies of the ESO/SRC
and a magnification of 50. The surveys are divided into the Hydra to
Puppis region (D$_1$: Salem \& Kraan-Korteweg, in prep.), the
Hydra/Antlia Supercluster region (D$_2$: Kraan-Korteweg 2000), the
Crux region (D$_3$: Woudt 1998, Woudt \& Kraan-Korteweg 2000a), the GA
region (D$_4$: Woudt 1998, Woudt \& Kraan-Korteweg 2000b), and the
Scorpius region (D$_5$: Fairall \& Kraan-Korteweg, 2000). Slightly
over 17\,000 galaxies were identified in these regions of which $\sim
97\%$ were previously unknown. Folding the galaxy distribution with
extinction maps revealed various unknown clumpings, the most
impressive recognition being that the cluster A3627 at ($\ell,b,v) =
(325\deg,-7\deg,4848$~\kms) within the Great Attractor region (see
Fig.~\ref{D1_5}) would be the most prominent galaxy overdensity in the
southern sky were it not for the diminishing effects of the foreground
extinction (Kraan-Korteweg \etal 1996).

{\bf Region E:} 
Motivated by the chance discovery of
two clusters behind the Galactic bulge, \ie the Ophiuchus cluster
at $(\ell,b,v) = (0\fdg5, 9\fdg0, 8600$~\kms) by Johnston \etal (1981) and
Wakamatsu \& Malkan (1981), as well as the
Sagittarius cluster closeby in redshift space ($359\fdg8,
8\fdg0, 8400$~\kms), Wakamatsu and collaborators surveyed this region
in more detail. They performed a deep survey of six ESO/SRC fields centered 
on these clusters and a shallow survey from $16^h10^m < \alpha <17^h50^m,
-32\fdg5 < \delta < 0\deg$ to search for wall-like connections with
the Hercules cluster (region E). In the former region close to
4000 galaxies were charted with $D \ga 0.1$\,mm, revealing two
new clusters and 4 galaxy groups all at the same redshift
range (Wakamatsu \etal 1994, Hasegawa \etal 2000).

{\bf Region F}: 
In 1995, Hau \etal (1995) searched 12 red POSS plates at $\ell \sim
135\deg$ because of the increased likelyhood of detecting galaxies
along the Supergalactic Plane. They indeed identified a signficant
number of galaxies (N = 2575), though this relatively high number is
also due to the fact that this search includes higher latitude ($|b| \la
25\deg$) regions, hence lower extinction levels compared to the other
searches. To confirm the nature of these galaxy candidates, follow-up 
observations using various techniques were performed on a sample
of suspected nearby galaxies (Lahav \etal 1998).

A comparison of the surveyed regions (Fig.~\ref{cor}) with the ZOA as
outlined in Fig.~\ref{ait} clearly demonstrates that nearly the whole
ZOA has been covered by systematic deep optical galaxy searches. All
these searches have similar characteristics and reveal that galaxies
can easily be traced through obscuration layers of 3 magnitudes,
narrowing therewith the ZOA considerably. This is illustrated with
Fig.~\ref{D1_5} which shows an area of the sky centered on the
southern Milky Way with all the Lauberts galaxies larger than $D \ge
1\farcm3$ (diameter-coded as in Fig.~\ref{ait}) plus all the galaxies
with $D \ge 12 \arcsec$ from the deep optical galaxy searches by
Kraan-Korteweg and collaborators (D$_1$-D$_5$ in Fig.~\ref{cor}).
DIRBE extinction contours equivalent to $A_{B} = 1\fm0$ and $3\fm0$
are also drawn.

A few galaxies still are recognizable up to extinction levels of
$A_{B} = 5\fm0$ and a handful of very small galaxy candidates have
been found at even higher extinction levels. The latter ones most
likely indicate holes in the dust layer.  Overall, the mean number
density follows the dust distribution remarkably well. The contour
level of $A_{B} = 5\fm0$, for instance, is nearly indistinguishable
from the galaxy density contour at 0.5 galaxies per square degree.

Analyzing the galaxy density as a function of galaxy size, magnitude
and/or morphology in combination with the foreground extinction has
led to the identification of various important large-scale structures
and their approximate distances. In Fig.~\ref{D1_5}, for instance, the
most extreme overdensity is found at $(\ell,b) \sim
(325\deg,-7\deg$). It is at least a factor 10 denser compared to
regions at similar extinction levels. This galaxy excess is centered
on the cluster A3627, now recognized as the most massive cluster in
the nearby Universe (Kraan-Korteweg \etal 1996, see also
Sect.~\ref{GA}). To trace these structures in detail, an understanding
of the completeness of these searches is required and follow-up
observations must be obtained to map the large-scale structures in
redshift space (see Sect.~\ref{optred}).

\subsection{Completeness of optical galaxy searches}\label{optcomp}

In order to merge the various deep optical ZOA surveys with existing
galaxy catalogs, Kraan-Korteweg (2000) and Woudt (1998) have analyzed
the completeness of their ZOA galaxy catalogs -- the Hyd/Ant [D$_2$],
Crux [D$_3$] and GA [D$_4$] region -- as a function of the foreground
extinction.

By studying the apparent diameter distribution as a function of the
extinction $E(B-V)$ (Schlegel \etal 1998) as well as the location of
the flattening in the slope of the cumulative diameter curves $(\log
D) - (\log N)$ for various extinction intervals (\cf Fig.~5 and 6 in
Kraan-Korteweg 2000), they conclude that their optical ZOA surveys are
complete to an apparent diameter of ${D} = 14\arcsec$ -- where the
diameters correspond to an isophote of 24.5~mag/arcsec${^2}$ -- for
extinction levels less than $A_{B} = 3\fm0$.

How about the intrinsic diameters, \ie the diameters galaxies would
have if they were unobscured? A spiral galaxy seen through an
extinction of $A_{B} = 1\fm0$ will, for example, be reduced to $\sim
80\%$ of its unobscured size.  Only $\sim 22\%$ of a (spiral) galaxy's
original dimension is seen when it is observed through $A_{B} =
3\fm0$.  In 1990, Cameron derived analytical descriptions to correct
for the obscuration effects by artificially absorbing the intensity
profiles of unobscured galaxies. These corrections depend quite
strongly on morphological type due to the mean surface brightness 
and difference in brightness profiles between early-type
galaxies and spiral galaxies. Applying these corrections,
Kraan-Korteweg (2000) and Woudt (1998) found that at $A_{B} = 3\fm0$,
an obscured spiral or an elliptical galaxy at their {\it apparent}
completeness limit of ${D} = 14\arcsec$ would have an intrinsic
diameter of ${D^o} \sim 60\arcsec$, respectively ${D^o} \sim
50\arcsec$. At extinction levels higher than $A_{B} = 3\fm0$, an
elliptical galaxy with $D^o = 60\arcsec$ would appear smaller than the
completeness limit $D = 14\arcsec$ and might have gone unnoticed.
These optical galaxy catalogs should therefore be complete to $D^o \ge
60\arcsec$ for galaxies of all morphological types down to extinction
levels of $A_{B} \le 3\fm0$ with the possible exception of extremely
low-surface brightness galaxies. Only intrinsically very large and
bright galaxies -- particularly galaxies with high surface brightness
-- will be recovered in deeper extinction layers.  This completeness
limit could be confirmed by independently analyzing the diameter
vs. extinction and the cumulative diameter diagrams for
extinction-corrected diameters.

\begin{figure}[p]
\begin{center}
\hfil \psfig{file=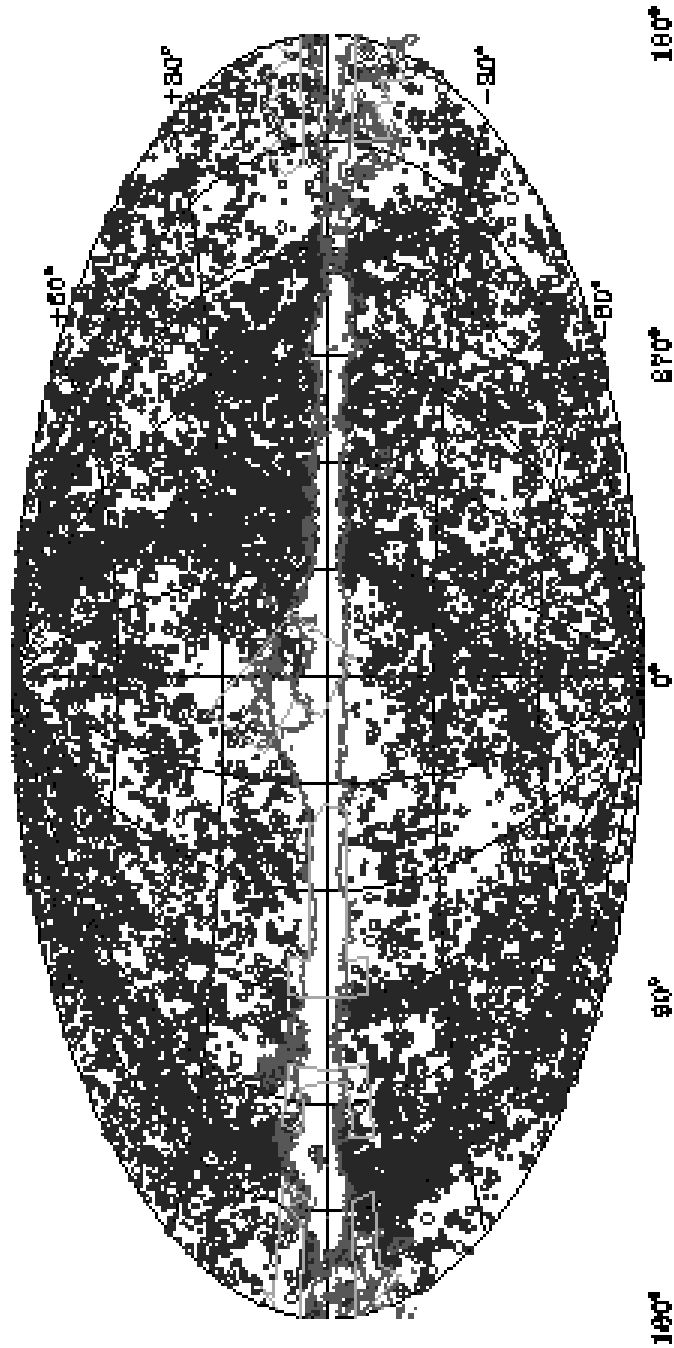,angle=90,height=18cm} \hfil
\caption
{Aitoff equal-area distribution of ESO, UGC, MCG galaxies with 
extinction-corrected diameters ${D^o} \ge 1\farcm3$, including 
galaxies identified in the optical ZOA galaxy searches for 
extinction-levels of $A_{B} \le 3\fm0$ (contour). The diameters are
coded as in Fig.~1. With the exception of the areas for
which either the positions of the galaxies or their diameters are
not yet available (demarcated areas), the ZOA could be reduced
considerably compared to Fig.~1.}
\label{aitc}
\end{center}
\end{figure}

One can thus supplement the ESO, UGC and MCG catalogs -- which are
complete to $D = 1\farcm3$ -- with galaxies from optical ZOA galaxy
searches that have ${D^o} \ge 1\farcm3$ and $A_{B} \le 3\fm0$. As the
completeness limit of the optical searches lies well above the ESO,
UGC and MGC catalogs, one can assume that the other similarly
performed optical galaxy searches in the ZOA should also be complete
for galaxies with extinction-corrected diameters ${D^o} \ge 1\farcm3$
to extinction levels of $A_{B} \le 3\fm0$.
In Fig.~\ref{aitc}, we have then taken the first step in arriving at
an improved whole-sky galaxy distribution with a reduced ZOA. In this
Aitoff projection, we plot all the UGC, ESO, MGC galaxies that
have {\it extinction-corrected} diameters ${D^o} \ge 1\farcm3$
(remember that galaxies adjacent to the optical galaxy search regions
are also affected by absorption though to a lesser extent: 
$A_{B} \le 1\fm0$), and added all the galaxies from the various optical
surveys with ${D^o} = 1\farcm3$ and $A_{B} \le 3\fm0$ for
which positions and diameters were available. The regions for which
these data are not yet available are marked in Fig.~\ref{aitc}. As
some searches were performed on older generation POSS I plates,
which are less deep compared to the second generation POSS II and
ESO/SERC plates, an additional correction was applied to those
diameters, \ie the same correction as for the UGC galaxies which also
are based on POSS I survey material (${D_{25} = 1.15 \cdot 
D_{POSS I}}$).

A comparison of Fig.~\ref{ait} with Fig.~\ref{aitc} demonstrates 
convincingly how the deep optical galaxy searches realize a 
considerable reduction of the ZOA: we can now trace the
large-scale structures in the nearby Universe to extinction levels
of $A_{B} = 3\fm0$. Inspection of Fig.~\ref{aitc} reveals that
the galaxy density enhancement in the GA region is even more
pronounced and a connection of the Perseus-Pisces chain across the
Milky Way at $\ell=165\deg$ more likely. Hence, these supplemented
whole-sky maps certainly should improve our understanding of the 
velocity flow fields and the total gravitational attraction on 
the Local Group.

\subsection{Redshift follow-ups of optical surveys} \label{optred}
The analysis of the galaxy density as a function the foreground
extinction revealed various large-scale structures in the ZOA. These
need to be mapped in redshift space. So far, the Perseus-Pisces
supercluster, the Puppis region, the Ophiuchus supercluster behind the
Galactic Bulge area, and the southern ZOA have been intensively
observed. The most prominent new galaxy structures revealed in this
way are summarized below. Their approximate positions (ordered in
Galactic latitude) are given as ($\ell, b, v$):

Based on a redshift survey of over 2500 galaxies using the
multi-spectrograph ``Flair'' on the UKST (AAO), Wakamatsu \etal (2000)
confirmed that the Ophiuchus cluster behind the Galactic bulge
($0\fdg5, 9\fdg5, 8500$\kms) is the central component of a
supercluster including two more clusters and four groups of
galaxies. There seems to be a wall-like structure connecting the
Ophiuchus cluster to the Hercules supercluster at 11000~\kms. This
wall runs orthogonal to the Great Wall.

At ($\ell, b$)$\sim$ ($33\deg, 5\deg-15\deg$), Marzke \etal
(1996) and Roman \etal (1998) found evidence for a nearby cluster
close to the Local Void at 1500~\kms, as well as a prominent cluster
behind the Local Void at 7500~\kms. The nearby cluster is
independently supported by data from blind \HI-surveys (see
Sect.~\ref{hi}).

The connection of the Perseus-Pisces supercluster across the ZOA to
the cluster A569, suspected by Focardi \etal (1984), was confirmed by
Chamaraux \etal (1990) and Pantoja \etal (1997).  The Perseus-Pisces
chain seems to fold back into the ZOA at higher redshifts at
($95\deg, -10\deg, 7500$~\kms), Marzke \etal (1996), Pantoja \etal
(1997).

In 1992, Kraan-Korteweg \& Huchtmeier uncovered a nearby
cluster in Puppis ($245\deg, 0\deg, 1500$~\kms) which was later shown by
Lahav \etal (1993) to contribute a non-negligible component to the
peculiar z-motion of the Local Group. Since then, this region
has been investigated in further detail. Chamaraux \etal (1999)
found further evidence for this cluster. It lies within
a long narrow filament (Masnou \& Chamaraux, in prep.) which 
extends from the Antlia to the Fornax cluster 
(see also Fig.~\ref{vslice}). 

Kraan-Korteweg \etal (1994a) presented evidence for a
continous filamentary structure extending over $30\deg$ on the sky
from the Hydra and Antlia clusters across the ZOA, intersecting the
Galactic Plane at ($280\deg, 0\deg, 3000$~\kms). At the same longitudes,
they noted significant clustering at $\sim$ 15000~\kms, indicative of
a connection between the Horologium and Shapley clusters a hundred
degrees apart in the sky.

Kraan-Korteweg \& Woudt (1993) uncovered a shallow but
extended supercluster in Vela at ($285\deg, 6\deg, 6000$~\kms).

Next to the massive cluster A3627 at the core of the Great
Attractor (clustering in the Great Attractor region is discussed in
the next section), Woudt (1998) discovered a cluster at ($306\deg,
6\deg,6200$) called the Cen-Crux cluster, and a more distant cluster,
the Ara cluster at ($329\deg, -9\deg, 15000$~\kms). The latter might be
connected to the Triangulum-Australis cluster.

\subsubsection{Clustering within the Great Attractor region}\label{GA}

Based on a deep optical galaxy search and subsequent redshift
follow-ups, Kraan-Korteweg \etal (1996) and Woudt (1998) have clearly
shown that the Norma cluster, A3627, at ($325\deg, -7\deg$, 4848\kms)
is the most massive galaxy cluster in the GA region known to date and
probably marks the previously unidentified but predicted density-peak
at the bottom of the potential well of the GA overdensity. The
prominence of this cluster has independently been confirmed by ROSAT
observations: the Norma cluster ranks as the 6$^{th}$ brightest X-ray
cluster in the sky (B\"ohringer \etal 1996). It is comparable in size,
richness and mass to the well-known Coma cluster.
Redshift-independent distance determinations (R$_{\rm C}$ and I$_{\rm
C}$ band \tfr\ analysis) of the Norma cluster have shown it to be at
rest with respect to the rest frame of the Cosmic Microwave Background
(Woudt 1998).

One cannot, however, exclude the possibility that other unknown rich
clusters reside in the GA region as the ZOA has not been fully unveiled
with optical searches.  Finding a hitherto uncharted, rich cluster of
galaxies at the heart of the GA would have serious implications for
our current understanding of this massive overdensity in the local
Universe.  Kraan-Korteweg \& Woudt (1999) found various indications
that PKS1343$-$601, the second brightest extragalactic radio source in
the southern sky ($f_{20cm} = 79$~Jy, McAdam 1991, and references
therein) might form the center of yet another highly obscured rich
cluster, particularly as it also shows significant X-ray emission
(Tashiro \etal 1998): extended diffuse hard X-ray emission at the
position of PKS1343$-$601 has been detected with ASCA. The radiation, 
kT = 3.9~keV, is far too large for it being associated with a
galactic halo surrounding the host galaxy, hence it might be
indicative of emission from a cluster -- if it is not due to the Inverse
Compton process.

At ($\ell, b) \sim (310\deg, 2\deg$), this radio galaxy lies behind an
obscuration layer of about 12 magnitudes of extinction in 
the B-band, as estimated from the DIRBE extinction maps (Schlegel 
\etal 1998). Its observed diameter of 28 arcsec in the Gunn-z filter 
(West \& Tarenghi 1989) translates into an extinction-corrected
diameter of 232 arcsec (following Cameron 1990). With a recession 
velocity of $v = 3872$~\kms\ (West \& Tarenghi 1989), this galaxy can 
be identified with a giant elliptical galaxy.
 
\begin{figure}[hb]
\begin{center}
\hfil \psfig{file=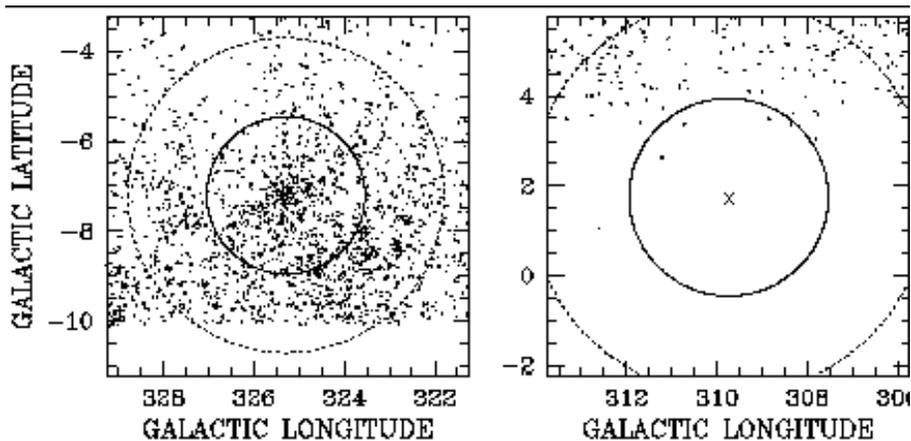,width=12cm} \hfil
\caption{A comparison of the rich A3627 cluster ($A_{B}\sim 1\fm5$)
and the suspected cluster centered on PKS1343$-$601 ($A_{B}\sim
12^{m}$), both in the GA region. Small dots are optically identified
galaxies. The inner circle marks the Abell radius $R_{\rm A}$ = 3
$h_{50}^{-1}$ Mpc.}
\label{pks1343}            
\end{center}
\end{figure}

Since PKS1343$-$601 is so heavily obscured, little data are available
to substantiate the existence of this prospective cluster.  In
Fig.~\ref{pks1343}, the A3627 cluster at a mean extinction $A_B =
1\fm5$ as seen in deep optical searches is compared to the prospective
PKS1343 cluster at ($309\fdg7, +1\fdg7, 3872$~\kms) with an extinction
of 12$^{\rm m}$.  One can clearly see that at the low Galactic
latitude of the suspected cluster PKS1343, the optical galaxy survey
could not retrieve the underlying galaxy distribution, especially not
within the Abell radius of the suspected cluster (the inner circle in
Fig.~\ref{pks1343}). To verify this cluster, other observational
approaches are necessary. We have imaged the prospective cluster
within its Abell radius in the near infrared (Woudt \etal in
progress). These observations will allow us to determine whether or
not PKS1343$-$601 is embedded in a centrally condensed overdensity of
galaxies.  Interestingly enough, deep \HI\ observations did uncover a
significant excess of galaxies at this position in velocity space (see
Sect.~\ref{MBdeep}) although a ``finger of God'', the characteristic
signature of a cluster in redshift space, is not seen.  Hence, the
Norma cluster A3627 remains the best candidate for the center of the
extended GA overdensity.

\subsection{The Sagittarius dwarf}
\label{Sag}
A remarkable discovery was made by Ibata \etal (1994). They found a
nearby `dwarf' galaxy with diameter of about 3~kpc. This new galaxy,
named the Sagittarius dwarf, is on the far side of the Galactic
center, about 25~kpc away from us, but well inside the Milky Way.
This galaxy is most probably undergoing some tidal disruption, before
being absorbed by the Milky Way.  An interesting feature of the
discovery of this galaxy is that it was based on velocities, not
direct detection on plates. In fact, the Sagittarius dwarf galaxy is
some twenty degrees from end to end, making it the largest structure
in the sky after the Milky Way itself. Nonetheless, since the new
galaxy lies directly behind the central bulge of the Milky Way, it
cannot be seen in direct images - even with hindsight - through the
dust and against the very much larger number of Galactic stars.

The Sagittarius dwarf galaxy provides an important clue to the
formation process of the Milky Way. Many popular models of galaxy
formation suggest that large galaxies are formed by a long process of
aggregation of many smaller galaxies, possibly with some merger
events being disruptive of the normal disk structure of spiral
galaxies. Such a process should still be common today, yet it had been
observed previously only in extremely rare cases. The Sagittarius dwarf
merger with the Milky Way provides the `smoking gun', showing that
such mergers do happen, they happen today, and they are not
destructive.

\subsection{Conclusions}

Deep optical galaxy searches have succesfully reduced the solid angle
of the ZOA by a factor of about $2-2.5$ down to extinction levels of
${A_B} = 3\fm0$ and have identified a number of important unknown
structures.  However, they fail in the most opaque part of the Milky
Way, the region encompassed by the ${A_B} = 3\fm0$ contour in
Fig.~\ref{aitc} -- a sufficiently large region to hide further
dynamically important galaxy densities. Here, systematic surveys
in other wavebands can be applied to reduce the current ZOA even
further. The success and status of these approaches are discussed in
the following sections.

\section{Near infrared surveys} \label{nir}
Observations in the near infrared (NIR) can provide important
complementary data to other surveys. With extinction decreasing as a
function of wavelength, NIR photons are up to 10 times less affected
by absorption compared to optical surveys -- an important aspect in the
search and study of galaxies behind the obscuration layer of the Milky
Way. The NIR is sensitive to early-type galaxies -- tracers of
massive groups and clusters -- which are missed in IRAS and \HI\
surveys (Sect.~\ref{fir} and ~\ref{hi}).  In addition, confusion with
Galactic objects is considerably lower compared to the FIR surveys.
Furthermore, because recent star formation contributes only little
to the NIR flux of galaxies (in contrast to optical and FIR emission),
NIR data give a better estimation of the stellar mass content of
galaxies. It is therefore well suited for the application of the \tfr\
either through pointed \HI\ observations of galaxies detected in the
NIR or a merging of detections from systematic blind \HI\ surveys with
NIR observations (Sect.~\ref{nirhi}).

\subsection{The NIR surveys DENIS and 2MASS}
Two systematic near infrared surveys are currently being performed:
DENIS, the DEep Near Infrared Southern Sky Survey, is imaging the
southern sky from $-88\deg < \delta < +2\deg$ in the \II\ (0.8\micron),
\J\ ($1.25\mu$m) and \K\ ($2.15\mu$m) bands. 2MASS, the 2 Micron All
Sky Survey, is covering the whole sky in the \J\ ($1.25\mu$m), \HH\
($1.65\mu$m) and \K\ ($2.15\mu$m) bands.  The mapping of the sky is
performed in declination strips, which are $30\deg$ in length and
12~arcmin wide for DENIS, and $6\deg \times 8\farcm5$ for
2MASS. Both the DENIS and 2MASS surveys are expected to complete their
observations by the end of 2000.  The main characteristics of the 2
surveys and their respective completeness limits for extended sources
are given in Table~\ref{nirprop} (Epchtein 1997, 1998, Skrutskie \etal
1997, Skrutskie 1998).

\begin{table}
\caption{Main characteristics of the DENIS and 2MASS surveys}
\label{nirprop}
\begin{tabular}{l|ccc|ccc}
\hline 
\vspace{-1mm} \\
& \multicolumn{3}{c|}{DENIS} & \multicolumn{3}{c}{2MASS} \\
Channel            &  \II\     & \J\       & \K\  &   \J\       & \HH\       & \K\        \\
\vspace{-1mm} \\
\hline 
\vspace{-1mm} \\
Central wavelength &  $0.8\mu$m   & $1.25\mu$m  & $2.15\mu$m  & $1.25\mu$m  & $1.65\mu$m & $2.15\mu$m \\   
Arrays             & 1024x1024   &  256x256   &  256x256   &  256x256   &  256x256  & 256x256    \\
Pixel size         & $1\farcs0$  & $3\farcs0$ & $3\farcs0$ & $2\farcs0$ & $2\farcs0$ & $2\farcs0$   \\
Integration time   &  9s         & 10s        & 10s        & 7.8s       & 7.8s       & 7.8s       \\
Completeness limit &&&&&\\
for extended sources & $16\fm5$  & $14\fm8$   & $12\fm0$   & $15\fm0$   & $14\fm2$   & $13\fm5$    \\ 
Number counts for the &&&&&\\
completeness limits &    50	 &  28        &    3       &  48     &
$\sim$40       & 24       \\
Extinction compared &&&&&&\\
to the optical $A_B$ &    0.45	 &  0.21      & 0.09       &  0.21     &  0.14       & 0.09       \\ 
\vspace{-1mm} \\
\hline 
\end{tabular}
\end{table}

Details and updates on completeness, data releases and
data access for DENIS and 2MASS can be found on the 
websites http://www-denis.iap.fr, and 
http://www.ipac.caltech.edu/2mass, respectively.

The DENIS completeness limits (total magnitudes) for highly reliable
automated galaxy extraction (determined away from the ZOA, \ie $|b| >
10\deg$) are $I = 16\fm5$, $J = 14\fm8$, $K_s = 12\fm0$ (Mamon
1998). The number counts per square degrees for these completeness
limits are 50, 28 and 3 respectively.  For the 2MASS, the completeness
limits are $J = 15\fm0$, $H = 14\fm2$, $K_s = 13\fm5$ (isophotal
magnitudes), with number counts of 48, $\sim$40 and 24. Because no \HH\
counts have been published we estimated their counts from low, mid and
high-density fields and the $J-H$ and $H-K$ color distribution (\cf
http://spider.ipac.caltech.edu/staff/jarrett/2mass). The numbers agree
well with the 2MASS predictions of the detection of 1 million galaxies
on the whole sky for their \K\ band completeness limit. In all
wavebands, except \II, the number counts are still imprecise due to
the low number statistics and the strong dependence on the
star crowding in the analyzed fields. Still, these numbers suffice to
reveal the promise of NIR surveys for probing the galaxy distribution
at very low Galactic latitudes.

\begin{figure} [t]
\hfil \psfig{file=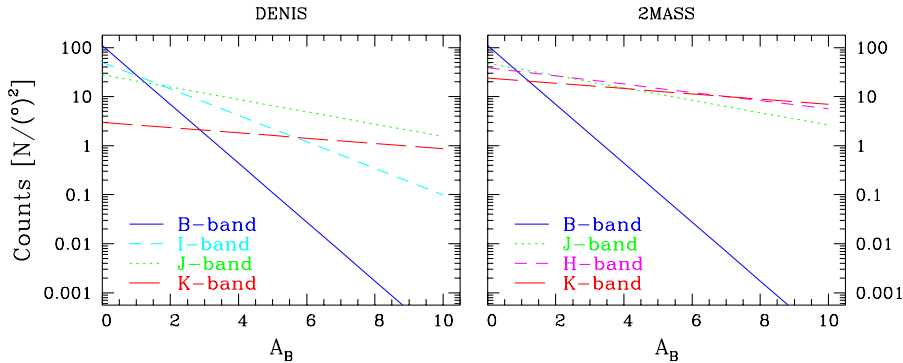,width=12cm} \hfil
\caption{Predicted \II, \J\ and \K\ galaxy counts for DENIS (left
 panel), and \J, \HH\ and \K\ counts for 2MASS (right panel) for their
respective galaxy completeness limits as a function of the absorption 
in the \B\ band. For comparison both panels also show the
B counts of an optical galaxy sample extracted from sky surveys.
}
\label{nircts}
\end{figure}

The decrease in number counts is considerably slower in the \II, \J,
\HH\ and \K\ bands compared to the optical because extinction is
only 45\%, 21\%, 14\% and 9\% compared to the \B\
band. This dependence makes NIR surveys very powerful at low Galactic
latitudes even though they are not as deep as the POSS and ESO/SERC
sky surveys.  As illustrated in Fig.~\ref{nircts}, the galaxy density
in the \B\ band in unobscured regions is 110 galaxies per square
degree for the completeness limit of $B_J\le19\fm0$ (Gardner \etal
1996). But the counts in the blue decrease rapidly with increasing
obscuration: $N(A_{B}) \simeq 110 \times {\rm dex}
(0.6\,[-A_{B}])\,$deg$^{-2}$. The decrease in detectable galaxies due
to extinction is slower in the NIR, and the counts of the shallower
NIR surveys overtake the optical counts at extinction levels of $A_B
\ga 1-3^m$.  The location of the reversal in efficiency is
particularly opportune because the NIR surveys become more efficient
where deep optical galaxy searches become incomplete, \ie at $A_B \ga
3\fm0$ (see Sect.~\ref{optcomp}).

The above predictions do not take into account any dependence on
morphological type, surface brightness, intrinsic color, orientation
and crowding, which may lower the counts of actually detectable
galaxies counts. In practice, \B\ was found to be superior for
identification of galaxies on DENIS images to extinction levels of at
least $A_B = 2\fm0$. And even though 2MASS appears more powerful in
Fig.~\ref{nircts} for ZOA research compared to DENIS, the higher
sensitivity of 2MASS also results in higher star densities at low
latitudes, making galaxy identifications more difficult. This problem
becomes apparent from, for instance, Fig.~28 and Fig.~29a in the 'List
of Figures' accessible from\\
http://spider.ipac.caltech.edu/staff/jarrett/2mass/3chan/basic/paper\_\,I.html
\\which show galaxy images in \J, \HH\ and \K\ in the ZOA at extinction
levels which are not yet very severe ($0\fm8$ and $2\fm6$ in the
optical).

\subsection{Pilot studies with DENIS data in the ZOA}

To compare the above predictions with real data, Schr\"oder \etal
(1997, 1999) and Kraan-Korteweg \etal (1998b) examined the efficiency of
uncovering galaxies at high extinctions using DENIS images. The
analyzed regions include the rich cluster A3627 ($\ell,b) =
(325\fdg3,-7\fdg2)$ at the heart of the GA (Norma) supercluster, as
well as its suspected extension across the Galactic Plane.


Three high-quality DENIS strips cross the cluster A3627.
66 images on these strips lie within the Abell-radius ($R_A =
1\fdg75$) and were inspected by eye (Kraan-Korteweg \etal 1998a).  This
covers about one-eighth of the cluster area.  The extinction over the
regarded cluster area varies as $1\fm2 \le A_B \le 2\fm0$.

On these 66 images, 151 galaxies had previously been identified in the
deep optical ZOA galaxy search (Woudt \& Kraan-Korteweg 2000b). Of
these, 122 were recovered in the \II, 100 in the \J, and 74 in the \K\
band. Most of the galaxies not re-discovered in \K\ are low surface
brightness spiral galaxies.

Surprisingly, the \J\ band provides better galaxy detection than the
\II\ band.  In the latter, the severe star crowding makes
identification of faint galaxies very difficult. At these extinction
levels, the optical survey does remain the most efficient in {\it
identifying} obscured galaxies.


The search for more obscured galaxies was made in the region $320\deg
\le \ell \le 325\deg$ and $|b|\le 5\deg$, \ie the suspected crossing
of the GA.  Of the 1800 images in that area, 385 of the then available
DENIS images were inspected by eye (308 in \K).  37 galaxies at higher
latitudes were known from the optical survey.  28 of these could be
re-identified in \II, 26 in \J, and 14 in the \K\ band. In addition,
15 new galaxies were found in \II\ and \J, 11 of which also appear in
the \K\ band. The ratios of galaxies found in \II\ compared to \B, and
of \K\ compared to \II\, are higher than in the A3627 cluster. This is
due to the higher obscuration level (starting with $A_B \simeq 2\fm3
-3\fm1$ at the high-latitude border).

On average, about 3.5 galaxies per square degree were found in the
\II\ band. This roughly agrees with the predictions of
Fig.~\ref{nircts}. Because of star crowding, we do not expect to find
galaxies below latitudes of $b \simeq 1\deg-2\deg$ in this longitude
range (Mamon 1994). Low-latitude images substantiate this -- the
images are nearly fully covered with stars.  Indeed, the lowest
Galactic latitude galaxies were found at $b \simeq 1\fdg2$ and $A_B
\simeq 11^{\rm m}$ (in \J\ and \K\ only).

\begin{figure} [t]
\hfil \psfig{file=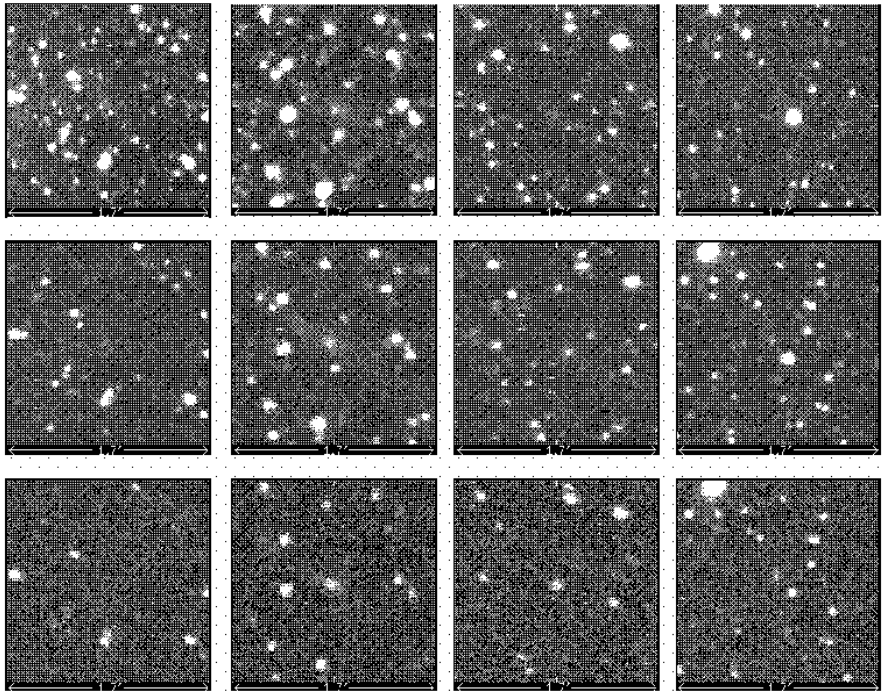,width=12cm} \hfil
\caption{DENIS survey images (before bad pixel filtering) of four
galaxies found in the deepest extinction layer of the Milky Way; 
the \II\  band image is at the top, \J\ in the middle and \K\ at the  bottom.} 
\label{nirbsex}
\end{figure}

Figure~\ref{nirbsex} shows a few characteristic examples of highly
obscured galaxies found in the DENIS blind search. \II\ band images
are at the top, \J\ in the middle and \K\ at the bottom. The first
galaxy located at $(l,b) = (324\fdg6,-4\fdg5$) is viewed through an
extinction layer of $A_B = 2\fm0$ according to the DIRBE extinction
maps (Schlegel \etal 1998). It is barely visible in the \J\ band. The
next galaxy at $(l,b) = (324\fdg7,-3\fdg5$) is subject to heavier
extinction ($A_B = 2\fm7$), and indeed easier to recognize in the
NIR. It is most distinct in the \J\ band. The third galaxy at even
higher extinction $(l,b,A_B) = (320\fdg1,+2\fdg5,5\fm7$) is -- in
agreement with the prediction of Fig.~\ref{nircts} -- not visible in
the \B\ band. Neither is the fourth galaxy at $b=+1\fdg9$ and $A_B =
9\fm6$: this galaxy can not be seen in \II\ band either and is very
faint only in \J\ and \K.


The conclusions from this pilot study are that at {\it intermediate
latitudes and extinction} ($|b| \ga 5\deg$, $A_B \la 4-5^{\rm m}$)
optical surveys are superior for identifying galaxies. But despite the
extinction and the star crowding at these latitudes, \II , \J\ and \K\
photometry from the survey data could be performed successfully at
these low latitudes. The NIR data (magnitudes, colors) of these
galaxies can therefore add important data in the analysis of these
obscured galaxies.  They led, for instance, to the preliminary
$I_c^o$, $J^o$ and $K_s^o$ galaxy luminosity functions in A3627
(Fig.~2 in Kraan-Korteweg \etal 1998a).

At the {\it lowest latitudes and at high extinction} ($|b| \la 5\deg$ and
$A_B \ga 4-5^{\rm m}$), the search for `invisible' obscured galaxies
on existing DENIS-images imply that NIR-surveys can trace galaxies
down to about $|b| \ga 1\deg - 1\fdg5$. The \J\ band was found to be
optimal for identifying galaxies up to $A_B \simeq 7^{\rm m}$.  NIR
surveys can hence further reduce the width of the ZOA.

\subsection{Systematic exploitation of NIR surveys and redshift
follow-ups}
\label{nirhi}
At the Observatoire de Lyon (France), DENIS images are being processed
routinely as they come out of the pipeline (cf. Vauglin \etal 1999 for a
catalog of over 20000 galaxies found on the \II\ band images in 25\%
of the southern sky).  Concerning the ZOA, \J\ and \K\ band DENIS
images are systematically being inspected by eye for galaxies below
$|b| \le 15\deg$ since March 1997.  This so far has led to the
detection of 1500 extended objects most of which were unknown
previously.  For these galaxies, magnitudes, isophotal diameters, axis
ratios, position angles and a rough estimate of the morphological type
are recorded.  This catalog will build the basis for systematic \HI\
line redshift follow-ups with the goal of applying the \tfr\ 
to determine the peculiar velocity field. Using a morphological
classification based upon a concentration parameter (Theureau \etal
1997), they want to obtain a complete sample of late-type inclined
spirals within a volume of $v < 10000$~\kms.

A similar project, but starting out from an \HI-selected sample, is
being pursued by cross-identifying all galaxies detected in the the
systematic deep blind \HI\ survey (Sect.~\ref{MB}) in the ZOA ($|b|
\le 5\fdg5$) on NIR images (Schr\"oder \etal in prep.) -- also with
the aim of determining the density field from the peculiar velocity
field through the \tfr.

A collaboration involving UMASS, IPAC and the CfA is working on a
redshift survey (the 2MASS Redshift Survey, 2MRS) of the whole sky that
will contain $\sim 250000$ galaxies to a limiting magnitude of \K\ $ =
13\fm5$. The first phase will be directed at galaxies with \K\ $ \le
12\fm2$. This survey will include the ZOA and although the
shallow redshift survey so far is only 25\% complete, it already shows
considerable structure through the Galactic Plane
out to high velocities (http://cfa-www.harvard.edu/\~\,huchra/2mass/).

Two spectroscopic follow-ups of DENIS and 2MASS galaxies are planned
on the 6dF robotic multi-object spectroscopic unit, currently under
construction at the AAO: a redshift survey of roughly 120000 
NIR-selected galaxies for which a total of 300 nights are guaranteed
between 2001 and 2003 and a peculiar velocity survey of roughly 12000
early-type galaxies. The latter will be complemented by an \HI\
peculiar velocity survey of over 5000 inclined spirals (Mamon 1998,
1999). In this project, however, only selected areas in the ZOA
will be probed.

\subsection{Conclusions}

First results from NIR data are very promising for ZOA research -- and
complementary to other approaches. NIR surveys become more efficient
in revealing galaxies at extinction levels where deep optical searches
become increasingly incomplete, i.e. at $A_B \simeq 3^{\rm m}$. It was
found that galaxies can be traced to Galactic latitudes of $|b| \ga 1
- 1\fdg5$.

A independent advantage of the NIR surveys is the fact that the NIR
colors, in particular of early-type galaxies, might help in the
calibration of the DIRBE extinction maps at low Galactic latitudes
(see \eg Fig.~5 in Schr\"oder \etal 1997).

The NIR surveys are particularly useful for the mapping of massive
early-type galaxies -- tracers of density peaks in the mass
distribution -- as these can not be detected with any of the
techniques that are efficient in tracing the spiral population 
in more opaque regions (Sect.~\ref{fir} and \ref{hi}).

Nevertheless, NIR surveys are also important with regard to the blue
and low surface-brightness spiral galaxies because a significant
fraction of them are detectable in the near infrared. This is
confirmed, for instance, with the serendipitous discovery in the ZOA
of a large, nearby ($v_{LSR} = 750$~\kms) edge-on spiral galaxy by
2MASS (Hurt \etal 1999): with an extension in the \K\ band of
5~arcmin, this large galaxy is -- not unexpectedly for its extinction
of $A_B = 6\fm6$ at the position of $(\ell,b) = (236\fdg8,-1\fdg8)$ --
not seen in the optical (Saito \etal 1991).  A more systematic
analysis of low-latitude spiral galaxies in the NIR was undertaken by
Schr\"oder \etal (priv. comm): about 80\% of the spiral galaxies
detected in the shallow systematic \HI\ survey performed with the
Parkes Multibeam Receiver (Sect.~\ref{MBsh}) could be reidentified on
\II\ band DENIS images, a few more in \J\ and \K\ only. The above nearby
NIR galaxy with a flux of 33.7 Jy~\kms\ in the 21~cm line was
discovered independently in this survey (HIZSS012, Henning \etal
2000).

The overlap of galaxies found in NIR and \HI\ surveys is important.
With the combination of \HI\ data and NIR data one can study the
peculiar velocity field via the NIR \tfr\ ``in the ZOA'' compared to
earlier interpolations of data ``adjacent to the ZOA''. This will 
provide important new input for density field reconstructions in the
ZOA (Sect.~\ref{recon}).

\section{Far infrared surveys}\label{fir}
In 1983, the Infrared Astronomical Satellite IRAS surveyed 96\% of the
whole sky in the far infrared bands at 12, 25, 60 and 100~$\micron$,
resulting in a catalog of 250\,000 point sources, i.e. the IRAS Point
Source Catalogue (IRAS PSC; Joint IRAS Science Working Group
1988). The latter has been used extensively to quantify extragalactic
large-scale structures.  The identification of the galaxies from the
IRAS data base is quite different compared to the optical: only the
fluxes at the 4 far infrared (FIR) IRAS passbands are available but no
images, and the identification of galaxies is strictly based on the
flux ratios. For instance, Yamada \etal (1993) used the
criteria: {\bf 1.}  $f_{60} > 0.6$Jy, {\bf 2.} $f^2_{60} >
f_{12} f_{25}$, {\bf 3.} $0.8 < f_{100}/f_{60} < 5.0$, to select
galaxy candidates from the IRAS PSC.

With these flux and color criteria, mainly normal spiral galaxies
and starburst galaxies are identified. Hardly any dwarf galaxies enter
the IRAS galaxy sample, nor the dustless elliptical galaxies, as
they do not radiate in the far infrared. The upper cut-off in the third
criterion is imposed to minimize the contamination with cool cirrus
sources and young stellar objects within our Galaxy. This, however, 
also makes the IRAS surveys less complete for nearby galaxies
(\eg Woudt 1998, Kraan-Korteweg 2000).

The advantage of using IRAS data for large-scale structure studies is
its homogeneous sky coverage (all data from one instrument) and the
negligible effect of the extinction on the flux at these long
wavelengths.  Even so, it remains difficult to probe the inner part of
the ZOA with IRAS data because of cirrus, high source counts of
Galactic objects in the Galaxy, and confusion with these objects --
most of them have the same IRAS characteristics as external
galaxies. The difficulty in obtaining unambiguous galaxy
identifications at these latitudes was demonstrated by Lu \etal (1990),
who found that the detection rate of IRAS galaxy candidates decreases
strongly as a function of Galactic latitude (from $|b| = 16\deg$ to
$|b| = 2\deg$). This can only be explained by the increase in faulty
IRAS galaxy identifications. Yamada \etal (1993) also found a dramatic
and unrealistic increase in possible galaxies close to the Galactic
Plane in their systematic IRAS galaxy survey of the southern Milky Way
($|b| \le 15\deg$).

\begin{figure}[t]
\begin{center}
\hfil \psfig{file=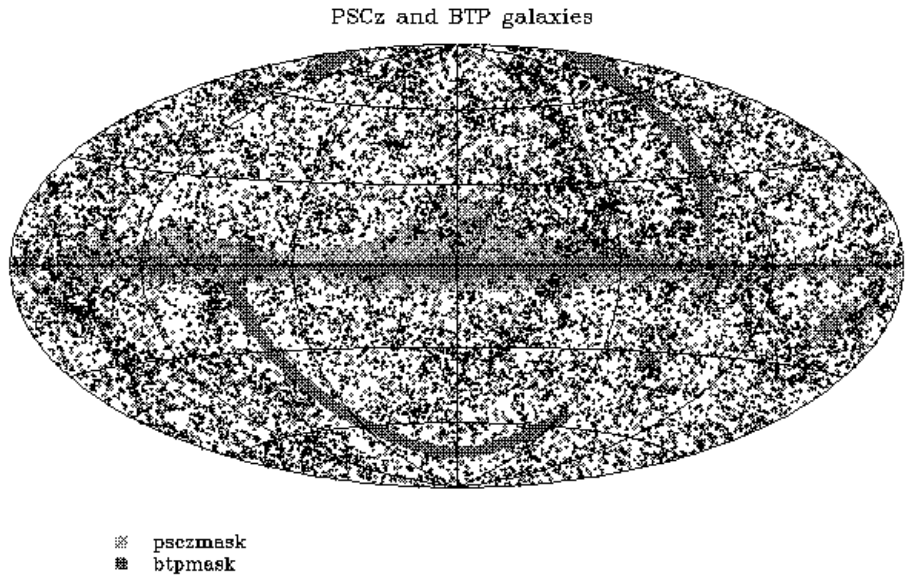,angle=-90,width=12cm} \hfil
\caption
{The PSCz and BTP IRAS galaxy catalogs centered on the Galaxy (same
orientation as previous projections) with the
PSCz incompleteness mask (light-grey) and the BTP incompleteness mask
(dark-grey). Note the dramatic reduction of the incompleteness around
the Galactic Equator due to the BTP survey.}
\label{BTP}
\end{center}
\end{figure}

So, despite the various advantages given with IRAS data, the sky
coverage in which reliable IRAS galaxy identifications can be made
(84\%) provides only a slight improvement over optical galaxy catalogs
(compare \eg the light-grey mask in Fig.~\ref{BTP} with the optical
ZOA-contour as displayed in Fig.~\ref{ait}). In addition to that, the
density enhancements are very weak in IRAS galaxy samples because (a)
the IRAS luminosity function is very broad, which results in a more
diluted distribution since a larger fraction of distant galaxies will
enter a flux-limited sample compared to an optical galaxy sample, and
(b) IRAS is insensitive to elliptical galaxies, which reside mainly in
galaxy clusters, and mark the peaks in the mass density distribution
of the Universe. This is quite apparent when comparing the IRAS
galaxy distribution (Fig.~\ref{BTP}) with the optical galaxy
distribution (Fig.~\ref{ait} and Fig.~\ref{aitc}).

Nevertheless, dedicated searches for large-scale clustering within the
whole ZOA ($|b| \le 15\deg$) have been made by various Japanese
collaborations (see Takata \etal 1996 for a summary).  They used
IRAS color criteria to select galaxy candidates which were
subsequently verified through visual examination on sky surveys, such
as the POSS for the northern hemisphere and the ESO/SRC for the
southern sky. Because of their verification procedure, this data-set
suffers, however, from the same limitations in highly obscured regions as
optical surveys.

Based on redshift follow-ups of these ZOA IRAS galaxy samples, they
established various filamentary features and connections across the
ZOA. Most coincide with the structures uncovered in optical work. In
the northern Milky Way, both crossings of the Perseus-Pisces arms into
the ZOA are very prominent -- considerably stronger in the FIR than at
optical wavelengths -- and they furthermore identified a new structure:
the Cygnus-Lyra filament at ($60\deg-90\deg, 0\deg, 4000$\kms). Across the
southern Milky Way they confirmed the three general concentrations of
galaxies around Puppis ($\ell = 245\deg$), the Hydra--Antlia extension
($\ell = 280\deg$; Kraan-Korteweg \etal 1995) and the Centauraus
Wall ($\ell = 315\deg$). However, the cluster A3627 is not seen, nor is
the Great Attractor very prominent compared to the optical or to
the POTENT reconstructions described in Sect.~\ref{recon}.

Besides the search for the continuity of structures across the
Galactic Plane, the IRAS galaxy samples have been widely used for the
determination of the peculiar motion of the Local Group, as well as
the reconstructions of large-scale structure across the Galactic Plane
(see Sect.~\ref{recon}). These analyses have been performed on the
two-dimensional IRAS galaxy distribution and, in recent years, as well
as on their distribution in redshift space through the availability of
redshift surveys for progressively deeper IRAS galaxy samples, \ie\
2658 galaxies to f$_{60\mu m} = 1.9$~Jy (Strauss \etal 1992), 5321
galaxies to f$_{60\mu m} = 1.2$~Jy (Fisher \etal 1995), and lately the
PSCz catalog of 15411 galaxies complete to f$_{60\mu m} = 0.6$~Jy with
84\% sky coverage and a depth of 20000~\kms\ (Saunders \etal 2000b).

The PSCz is deep enough to test the convergence of the dipole. The most
recent analysis of the IRAS PSCz dipole (Schmoldt \etal 1999; see
also Rowan-Robinson \etal 2000) finds that the acceleration vector
points about $15\deg$ away from the CMB dipole.  Assuming full
convergence at the sample boundary, about 2/3 of the measured
acceleration is generated within 4000~\kms. There is a non-negligible
contribution out to 14000~\kms, after which the acceleration amplitude
seems to have converged. 

Saunders and collaborators realized, however, that the 16\% of the sky
missing from the survey causes significant uncertainty, particularly
because of the location behind the Milky Way of many of the prominent
large-scale structures (superclusters as well as voids). In 1994, they
therefore started a longterm program to increase the sky coverage of
the PSCz. Optimizing their color criteria to minimize contamination by
Galactic sources ($f_{60} / f_{25} > 2$, $f_{60} / f_{12} > 4$, and
$1.0 < f_{100}/f_{60} < 5.0$), they extracted a further 3500 IRAS
galaxy candidates at lower Galactic latitudes (light-grey area of
Fig.~\ref{BTP}). Taking $K'$ band snapshots of all the galaxy
candidates of their `Behind The Plane' [BTP] survey, they could add a
thousand galaxies to the PSCz sample and reduce the coverage gap to a
mere 7\% (dark-grey area).

The resulting sky map of 16,400 galaxies (PSCz plus BTP) is shown in
Fig.~\ref{BTP} (from Saunders \etal 2000a). The BTP survey has reduced
the ``IRAS ZOA'' dramatically.  Some incompleteness remains towards
the Galactic Center, but large-scale structures can easily be
identified across most of the Galactic Plane. In the Great Attractor
region, the galaxies can be traced (for the first time with IRAS data)
to the rich cluster A3627 -- the suspected core of the GA
(Kraan-Korteweg \etal 1996). The BTP collaboration is currently
working hard on obtaining redshifts for these new and heavily obscured
galaxies and exciting new results on large-scale structure across the
Milky Way and dipole determinations can be expected in the near
future.

\section{HI surveys} \label{hi}
In the regions of the highest obscuration and infrared confusion, the
Galaxy is fully transparent to the 21cm line radiation of neutral
hydrogen. \HI-rich galaxies can readily be found at lowest latitudes
through the detection of their redshifted 21cm emission, though
early-type galaxies -- tracers of massive groups and clusters -- are
gas-poor and will not be identified in these surveys. Furthermore,
low-velocity extragalactic sources (blue- and red-shifted) within the
strong Galactic \HI\ emission will be missed, and galaxies close to
radio continuum sources may also be missed because of baseline ripples.

The advantage of blind \HI\ surveys in the Milky Way is not only the
transparency of the 21cm radiation to the thickest dust layers: with
the detection of an \HI\ signal, the redshift and rotational
properties of an external galaxy are immediately known, providing
insight not only on its location in redshift space but also on the
intrinsic properties of such obscured galaxies. The rotational
velocity can furthermore be used in combination with \eg NIR
photometry (Sect.~\ref{nirhi}) to determine the distribution in real
space from \tfr\ distances and the density field independent of
interpolations across the Milky Way from the peculiar velocity field.

Until recently, radio receivers were not sensitive and efficient 
enough to attempt large systematic surveys of the ZOA. But
in a pilot survey, Kerr \& Henning (1987) pointed the late 
300-ft telescope of Green Bank to 1900 locations in the ZOA (1.5\%
coverage) and detected 19 previously unknown spiral galaxies, 
proving therewith the effectiveness of this approach.

Since then two systematic blind \HI\ searches for galaxies behind the
Milky Way were initiated. The first -- the Dwingeloo Obscured Galaxies
Survey (DOGS) -- used the 25~m Dwingeloo radio to survey the whole
northern Galactic Plane for galaxies out to 4000~\kms\ (\cf
Kraan-Korteweg \etal 1994b, Henning \etal 1998, Rivers \etal
1999). Although the sensitivity was fairly low (40~mJy for a 1~hr
integration), the advantage of the small telescope aperture is rapid
areal coverage.

A more sensitive survey, probing a considerably larger volume (out to
12700~\kms), is being performed for the southern Milky Way with the 64~m
radio telescope of Parkes (Kraan-Korteweg \etal 1998a, Staveley-Smith
\etal 1998, Henning \etal 1999, 2000). A Multibeam (MB) receiver with
13 beams in the focal plane array (Staveley-Smith 1996) was
specifically constructed to efficiently search for galaxies not
identified in optical surveys because of low optical surface
brightness or high optical extinction.

In the following, the observing techniques of these two surveys as
well as the first results will be discussed.

\subsection{The Dwingeloo Obscured Galaxies Survey (DOGS)}

Since 1994, the Dwingeloo 25~m radio telescope has been dedicated to a
systematic search for galaxies in the northern Zone of Avoidance
($30\deg \le \ell \le 220\deg$, $|b| \le 5\fdg25$). The last few
patches of the survey were completed early 1999 using the Westerbork
array in total power mode.  The 20~MHz bandwidth was tuned to cover
the velocity range $0 \le v_{\rm LSR} \le 4000$~\kms. Negative
velocities were excluded because the Leiden/Dwingeloo Galactic \HI\
survey (Hartmann 1994, Hartmann \& Burton 1997) had already covered the
velocity range $-450 \le v_{\rm LSR} \le 400$~\kms, albeit with higher
rms.

The 25~m Dwingeloo telescope has a half-power-beamwidth (HPBW) of
36~arcmin.  With a DAS-1000 channel autocorrelator spectrometer at the
telescope backend, the coverage over the 20~MHz bandwidth resulted in
a velocity resolution of 4~\kms. With this resolution even the
galaxies with the narrowest linewidth are covered by several channels. The
15000 survey points are ordered in a honeycomb pattern with a grid
spacing of $0\fdg4$. Galaxies are generally detected in various
adjacent pointings, facilitating a more accurate determination of
their positions through interpolations. Each DOGS observation
consisted of a sequence of 5 contiguous pointings at constant Galactic
latitude. From this, 5 On-Off pairs were created in such a way that a
real galaxy will appear once as a positive and once as a negative
signal in two independent scans. The rms noise per channel typically
was $\sigma_{ch}=40$~mJy for a 1 hr integration (12 x 5min).

Because of the duration of the project (15000 hours not including
overhead and downtime) the strategy was to first conduct a fast search
of 5min integrations (${\rm rms}=175$~mJy) to uncover possible massive nearby
galaxies whose effect might yield important clues to the dynamics of
the Local Group. In the following, the results from the Dwingeloo
shallow and deep surveys are discussed.
 
\subsubsection{The shallow Dwingeloo survey and the discovery of
Dwingeloo 1} \label{DOGSsh}
The shallow Dwingeloo search (${\rm rms}=175$~mJy) has been completed
in 1996 yielding five objects (Henning \etal 1998),
three of which were known previously. The most exciting discovery was
the barred spiral galaxy Dwingeloo 1 (Kraan-Korteweg \etal 1994b).

\begin{figure}[ht]
\begin{center}
\hfil \psfig{file=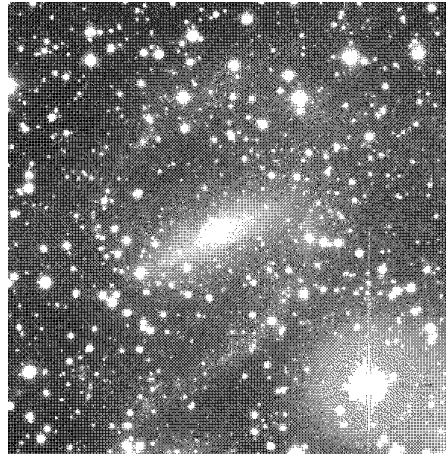,height=6cm} \hfil
\caption{Composite $V, R, I$-image of the Dwingeloo 1 galaxy at
$\ell=138\fdg5, b=-0\fdg1$. The displayed 484 x 484 pixels of
$0\farcs6$ cover an area of $4\farcm8$ x $4\farcm8$.  The large
diameter visible on this image is about $4\farcm2$.  Dwingeloo 1 has a
distinct bar, with 2 spiral arms that can be traced over nearly
$180\deg$. The morphology in this figure agrees with that of an SBb
galaxy.
}
\label{Dw1}            
\end{center}
\end{figure}

This galaxy candidate was detected early on in the survey through a
strong signal (peak intensity of 1.4~Jy) at the very low redshift of
$v_{\rm LSR} = 110$~\kms\ in the spectra of four neighbouring
pointings, suggestive of a galaxy of large angular extent.  The
optimized position of $(\ell,b)=(138\fdg5,-0\fdg1)$ coincided
with a very low surface brightness feature on the Palomar Sky Survey
plate of $2\farcm2$, detected earlier by Hau \etal (1995) in his
optical galaxy search of the northern Galactic/Supergalactic Plane
crossing (\cf Sect.~\ref{optsear}). Despite foreground obscuration of
about 6$^{\rm m}$ in the optical, follow-up observations in the $V$,
$R$ and $I$ band at the INT (La Palma) confirmed this galaxy candidate
as a barred, possibly grand-design spiral galaxy of type SBb of 4.2 x
4.2 arcmin (\cf Fig.~\ref{Dw1}).

Dwingeloo 1 has been the subject of many follow-up observations
(optical: Loan \etal 1996; Buta \& McCall 1999; HI-synthesis: Burton
\etal 1996; CO observations: Kuno \etal 1996; Li \etal 1996; Tilanus
\& Burton 1997; X-ray: Reynolds \etal 1997). To summarize, Dwingeloo 1 is a
barred spiral, with a rotation velocity of 130~\kms, implying a
dynamical mass of roughly one-third the mass of the Milky Way (within
the same fixed radius).  Its approximate distance of $\sim$ 3~Mpc
and angular location place it within the IC342/Maffei group of
galaxies.  The follow-up HI synthesis observations (Burton \etal 1996)
furthermore revealed a counterrotating dwarf companion, Dwingeloo
2. Since then various further dwarf galaxies have been discovered 
in this nearby galaxy group (next section).

\subsubsection{The deep Dwingeloo survey} \label{DOGSdeep}
Currently, 60\% of the deeper Dwingeloo survey (${\rm rms}=40$~mJy)
have been analyzed resulting in 36 detected galaxies, 23 of which were
previously unknown (Rivers \etal 1999). Five of the 36 sources were
originally identified in the shallow survey. Based on the survey
sensitivity, the registered number of galaxies is in agreement with
the Zwaan \etal (1997) HI mass function which predicts 50 to 100
detections for the full survey.

\begin{figure}[ht]
\begin{center}
\hfil \psfig{file=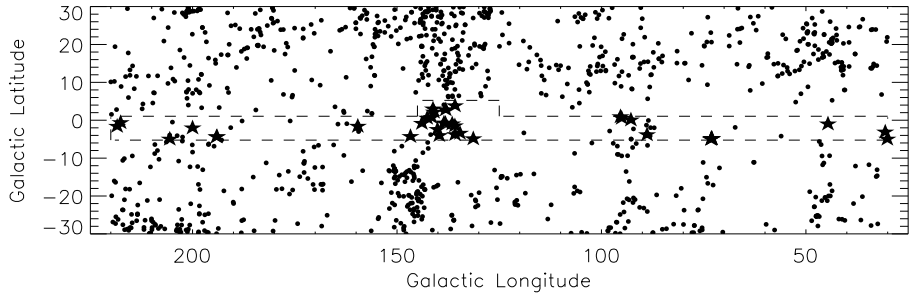,width=12cm} \hfil 
\vspace{0.4cm}
\caption{Spatial distribution of DOGS sources (indicated by $\star$)
combined with galaxies out to ${\rm v_{LSR}} \le 4000$~\kms\
($\bullet$) extracted from the LEDA database. The 60\% of
the analyzed part of the survey is outlined.}
\label{DOGSdet}
\end{center}
\end{figure}

The distribution of the Dwingeloo galaxies is shown in
Fig.~\ref{DOGSdet} together with other known galaxies out to
4000~\kms\ for visualization of connectivity of structures across the
Galactic Plane. Indeed, various known structures appear continuous
across the GP, although two galaxies were also found in
the Local Void ($\ell \sim 30\deg$). The latter two were
independently detected in the Parkes MB ZOA survey (Sect.~\ref{MBsh},
and Henning \etal 1999, 2000) and seem to be part of the by Roman \etal (1998)
recently discovered clustering at 1500\kms (Sect.~\ref{optred}).

11 galaxies were discovered in the Supergalactic Plane crossing region
($\ell \sim 137\fdg4$), five of which in the nearby IC342/Maffei
group: Dwingeloo~1, Maffei~2 from the shallow survey, plus three
dwarf members.  Further group members have been detected through
infrared photometry (McCall \& Buta 1995), and pointed observations
with the 100~m radiotelescope at Effelsberg of optically identified
dwarf candidates (Huchtmeier \etal 2000). The total number of obscured
IC342/Maffei group members has meanwhile grown to 19 members (for
recent updates on this group see Huchtmeier \etal 2000; Buta \& McCall
1999).  Photometric distances have been derived for 10 of these
galaxies (Karachentsev \& Tikhonov 1993, 1994; Karachentsev et
al. 1997), putting the group at a mere distance of $2.2\pm0.5$ Mpc.  At such
a close distance, this group might have played a significant role in
the dynamical history of the Local Group (McCall 1986, 1989; Zheng
\etal 1991; Valtonen \etal 1993; Peebles 1994; see Sect.~\ref{dyn}).

Suprisingly three dwarf galaxies were detected close to the nearby
isolated galaxy NGC 6946 at ($\ell,b,v_{\rm LSR}) = 
(95\fdg7,11\fdg7,46$~\kms). One of these had earlier been
cataloged as a compact high-velocity cloud (Wakker 1990).
Burton \etal (1999), in their search for compact isolated high-velocity
clouds in the Dwingeloo/Leiden Galactic \HI\ survey, discovered a
further member of this galaxy concentration.  Now, seven galaxies with
recessional velocities of $v_{\rm_{LSR}} \le 250$~\kms\ have been
identified within $15\deg$ of the galaxy NGC 6946. More might be
discovered as the DOGS data in this region have not yet been fully analyzed
(Fig.~\ref{DOGSdet}). The agglomeration of these various galaxies
might indicate a group or cloud of galaxies in the nearby
Universe.  As such it would be the only galaxy group in the nearby
Universe that is strongly offset (by $40\deg$) from the Supergalactic Plane
(Tammann \& Kraan-Korteweg 1978, Kraan-Korteweg 1979).

The most significant nearby, previously unknown galaxy identified with
DOGS was Dwingeloo 1.  Given the 80\% coverage of the survey region by
the shallow survey (Henning \etal 1998), chances are low that a
massive nearby spiral was missed, since nearby galaxies appear in many
adjacent pointings, all of which would have to be missed for the
galaxy to escape detection.  Thus, it is fairly unlikely that there 
exists another previously unidentified massive spiral galaxy in the area 
covered by the survey.

\subsection{The Parkes Multibeam ZOA blind HI survey} \label{MB}

In March 1997, the systematic blind \HI\ survey in the southern 
Milky Way ($212\deg \le \ell \le 36\deg$; $|b| \le 5\fdg5$) began
with the Multibeam receiver at the 64\,m Parkes telescope. The
instrument has 13 beams, each detecting orthogonal linear polarization.
The beams of ${\rm FWHP}=14\farcm4$ are arranged in a 
hexagonal grid in the focal plane (cf. Staveley-Smith \etal
1996), allowing rapid sampling of large areas. The average system 
temperature is about 20~K. 

The observations are being performed in driftscan mode. 23 contiguous
fields of length $\Delta\ell=8\deg$ have been defined.  Each field is
being surveyed along constant Galactic latitudes with latitude offsets
of 35~arcmin until the final width of $|b| \le 5\fdg5$ has been attained
(17 passages back and forth). The ultimate goal is to have 25 repetitions per
field where each repetition will furthermore be offset in latitude by
$\Delta b = 1\farcm5$ for homogeneous sampling. With an effective
integration time of 25 min/beam, a 3\,$\sigma$ detection limit of
25\,mJy is obtained. The survey covers the velocity range $-1200 \la v
\la 12700$~\kms\ with a channel spacing of 13.2~\kms\ per channel, and
will be sensitive to normal spiral galaxies well beyond the Great
Attractor region.  As a byproduct, the survey will produce a high
resolution integrated \HI\ column density map of the southern Milky
Way and a detailed catalog of high velocity clouds (\cf Putman \etal
1998).

So far, a shallow survey (next section) covering the whole southern
Milky Way, based on 2 out of the 25 foreseen driftscan passages, has
been analyzed (\cf Kraan-Korteweg \etal 1998a; Henning \etal 1999,
2000). A detailed study of the Great Attractor region ($308\deg \le
\ell \le 332\deg$) based on 4 scans has been made (Juraszek 1999;
Juraszek \etal 2000).  The first four full-sensitivity cubes are
available for that region as well (Sect.~\ref{MBdeep}).

\subsubsection{The Parkes ZOA MB shallow survey} \label{MBsh}
In the shallow survey, 110 galaxies were cataloged with peak \HI-flux
densities of $\ga$80~mJy (${\rm rms} = 15$~mJy after Hanning
smoothing). The detections show no dependence on Galactic latitude,
nor on the amount of foreground obscuration through which they have been
detected.  Though galaxies up to 6500~\kms\ were identified, most of
the detected galaxies (80\%) are quite local ($v<3500$\kms) due to the
(yet) low sensitivity. About one third of the detected galaxies have a
counterpart either in NED or in the deep optical surveys.

\begin{figure}[ht]
\begin{center}
\hfil \psfig{file=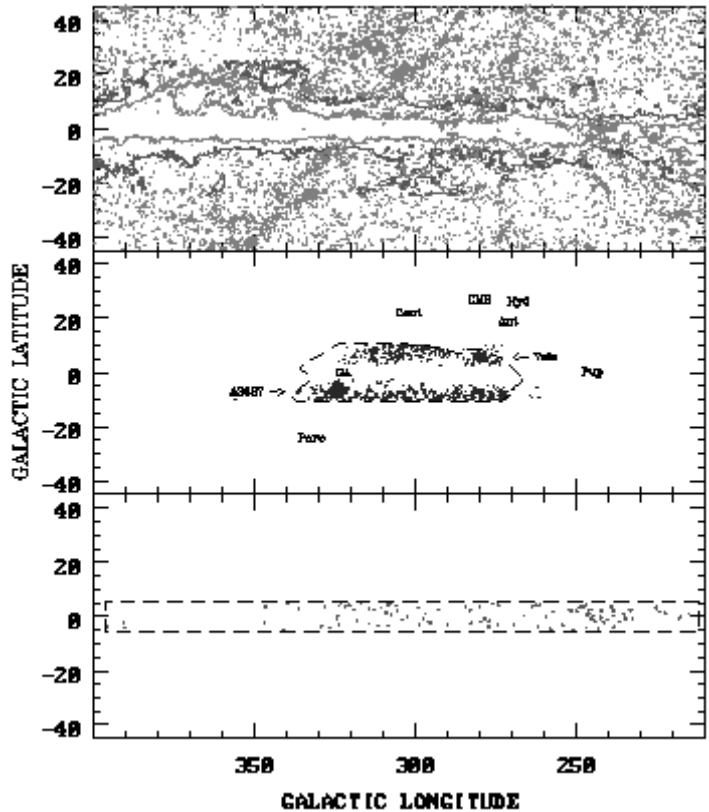,width=9.5cm} \hfil
\caption[]{Galaxies with $v<10000$~\kms.
Top panel: literature values (LEDA), superimposed
are extinction levels A$_B=1\fm0$ and $3\fm0$; middle panel:
follow-up redshifts (ESO, SAAO and Parkes) from the deep optical
ZOA survey with locations of clusters and dynamically important
structures; bottom panel: galaxies detected with the shallow Multibeam
ZOA survey. }
\label{vall}
\end{center}
\end{figure}

\begin{figure}[ht]
\begin{center}
\hfil \psfig{file=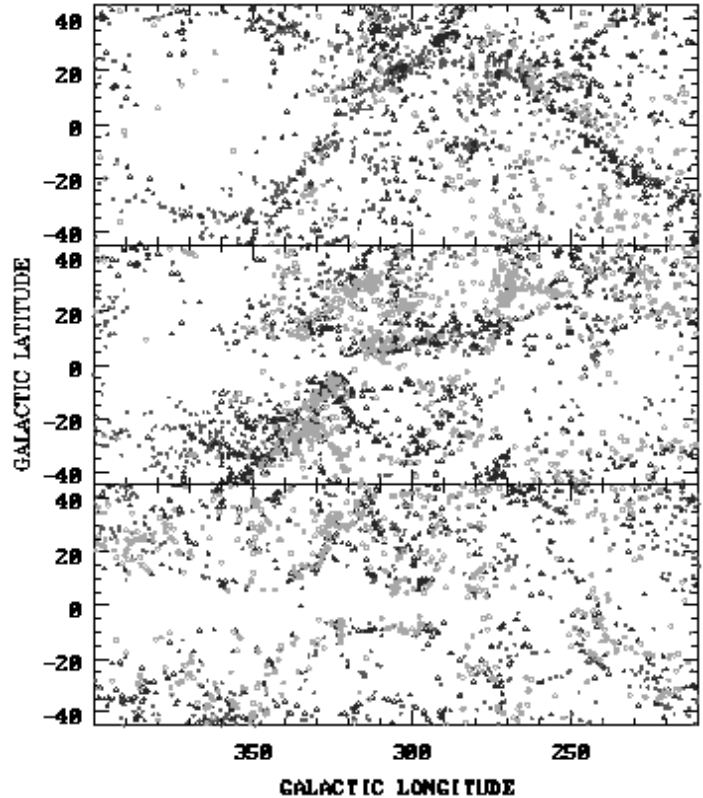,width=9.5cm} \hfil
\caption[]{Redshift slices from the data in Fig.~\ref{vall}: $500<v<3500$
(top), $3500<v<6500$ (middle), $6500<v<9500$~\kms\
(bottom). The open circles mark the nearest $\Delta v=1000$~\kms\ slice
in a panel, then triangles, then the filled dots the 2 more distant ones.}
\label{vslice}
\end{center}
\end{figure}

The distribution of the 110 \HI-detected galaxies is displayed in the
lower panel of Fig.~\ref{vall}. It demonstrates convincingly that
galaxies can be traced through the thickest extinction layers of the
Galactic Plane. The fact that hardly any galaxies are found behind the
Galactic bulge ($\ell=350\deg$ to $\ell=30\deg$) is due to local
structure: this is the region of the Local Void (see discussion below
and top panel of Fig.~\ref{vslice}).

For comparative purposes, the top panel of Fig.~\ref{vall} shows 
the distribution of all galaxies with known velocities v $\le
10000$~\kms\ to date (extracted from the LEDA database). Although 
this constitutes an uncontrolled sample it traces the main 
structures in the nearby Universe in a representative way. 
Note the increasing incompleteness for extinction levels of 
${A}_B \ga 1\fm0$ (outer contour) -- reflecting the
growing incompleteness of standard galaxy catalogs (see Sect.~\ref{opt}
and Fig.~\ref{ait}) -- and the almost complete lack of galaxy data for 
extinction levels ${A}_B \ga 3\fm0$ (inner contour).
The middle panel shows galaxies with  v$<$10000 \kms\ from 
the follow-up observations of the deep optical galaxy search by 
Kraan-Korteweg and collaborators (Sect.~\ref{optred}). Various new 
overdensities are apparent at low latitudes but the innermost part 
of our Galaxy remains obscured with this approach. Here, the 
blind \HI\ data (lower panel) finally can provide the missing link
for LSS studies. 

In Fig.~\ref{vslice}, the data of Fig.~\ref{vall} are combined in
redshift slices. The achieved sensitivity of the shallow MB \HI-survey
fills in structures all the way across the ZOA for the upper panel ($v
<$ 3500~\kms) for the first time. Note the continuity of the thin
filamentary sine-wave-like structure that dominates the whole southern
sky and crosses the Galactic equator twice. This structure snakes over
$\sim180\deg$ through the southern sky. Taking a mean distance of
$30{h^{-1}}$~Mpc, this implies a linear size of $\sim100{h^{-1}}$~Mpc,
with thickness of 'only' $\sim5{h^{-1}}$~Mpc or less. Various other
filaments spring forth from this dominant filament, always from a rich
group or small cluster at the junction of these interleaving
structures. This feature is very different from the thick, foamy Great
Wall-like structure, the GA, in the middle panel.

Also note the prominence of the Local Void which is very well
delineated in this presentation. No low redshift galaxies were found
within the Local Void. But three newly identified galaxies at
$\ell \sim 30\deg$ help to define the boundary of the Void.

The full-sensitivity ZOA MB-survey will fill in the LSS in the
more distant panels of Fig.~\ref{vslice}.  First results of
the full sensitivity survey have been obtained in the Great Attractor
region (next section).

Three nearby, very extended ($20\arcmin$ to $\ga 1\deg$) galaxies
were discovered in the shallow survey. Being likely candidates of
dynamically important galaxies, immediate follow-up observations were
initiated at the ATCA.  These objects did not turn out to be massive
perturbing monsters, however.  Two were seen to break up into \HI\
complexes and have unprecedented low \HI\ column densities 
(Staveley-Smith \etal 1998).  Systematic synthesis observations are
being performed to investigate the frequency of these interacting
and/or low \HI\ column density systems in this purely \HI-selected
sample.

\subsubsection{The Parkes ZOA MB deep survey} \label{MBdeep}
Four cubes centered on the Great Attractor region ($300\deg \ge \ell
\ge 332\deg$, $|b| \le 5\fdg5$) of the full-sensitivity survey have
been analyzed (Juraszek \etal 2000). 236 galaxies above the $3\sigma$ 
detection level of 25~mJy have been uncovered. 70\% of the detections
had no previous identification.

\begin{figure}[t]
\begin{center}
\hfil \psfig{file=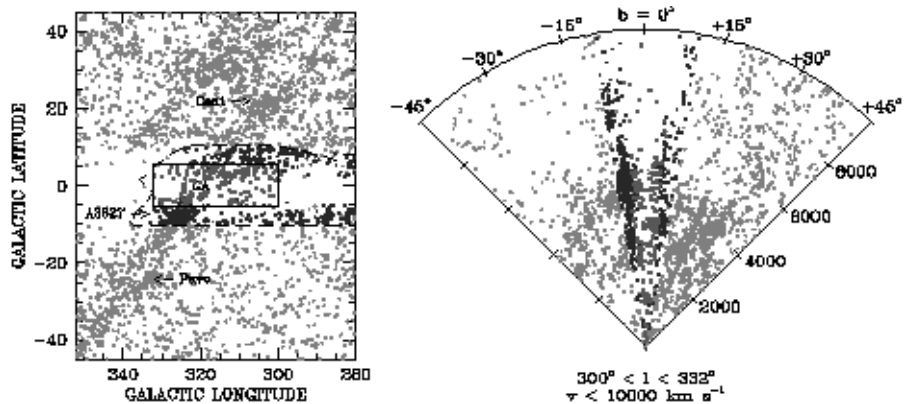,width=12cm} \hfil
\caption
{A sky distribution (left) and redshift cone (right) for galaxies with
$v<10000$~\kms\ in the GA region. Circles mark redshifts from the
literature (LEDA), squares redshifts from the optical galaxy search in
the Hyd/Ant-Crux-GA regions (outlined in the left panel) and crosses
detections in the full-sensitivity HI MB-ZOA survey (box).}
\label{MBGA}
\end{center}
\end{figure}

In the left panel of Fig.~\ref{MBGA}, a sky distribution centered on
the GA region displays all galaxies with redshifts ${v} \le
10000$~\kms. Next to redshifts from the literature (LEDA), redshifts
from the follow-up observations of Kraan-Korteweg and collaborators in
the Hyd/Ant-Crux-GA ZOA surveys (dashed area) are plotted. They clearly
reveal the prominence of the cluster A3627 at $(\ell,b,v) =
(325\deg,-7\deg,4848$~\kms; Kraan-Korteweg \etal 1996) close to the
core of the GA region at $(320\deg,0\deg,4500$~\kms), (Kolatt \etal
1995).  Adding now the new detections from the systematic blind \HI\
MB-ZOA survey (box), structures can be traced all the way across the
Milky Way. The new picture seems to support that the GA overdensity is
a ``great-wall'' like structure starting close to the Pavo cluster,
having its core at the A3627 cluster and then bending over towards
shorter longitudes across the ZOA.

This becomes even clearer in the right panel of Fig.~\ref{MBGA} where
the galaxies are displayed in a redshift cone out to ${v} \le
10000$~\kms\ for the longitude range $300\deg \le \ell \le 332\deg$
analyzed so far of the MB full-sensitivity data.  The A3627 cluster
is clearly the most massive galaxy cluster uncovered by the combined
surveys in the GA region and therefore the most likely candidate for
the previously unidentified but predicted density-peak at the bottom
of the potential well of the GA overdensity.

The new data do not unambigously confirm the existence of the
suspected further cluster around the bright elliptical radio galaxy
PKS1343$-$601 (Sect.~\ref{GA}). Although the MB data reveal an excess
of galaxies at this position in velocity space ($b=+2\deg,
v=4000$~\kms) a ``finger of God'' is not seen. It could be that many
central cluster galaxies are missed by the \HI\ observations because
spiral galaxies generally avoid the cores of clusters. The reality of
this possible cluster still remains a mystery. A first glimpse of the
$I$-band images obtained by Woudt \etal (in progress) reveal various
early-type galaxies. The forthcoming analysis should then
unambiguously settle the question whether another cluster forms part
of the GA overdensity.

\subsection{Conclusions}

The systematic probing of the galaxy distribution in the most opaque
parts of the ZOA with \HI\ surveys have proven very powerful. For the
first time LSS could be mapped without hindrance across the Milky Way
(Figs.~\ref{DOGSdet}, \ref{vslice} and \ref{MBGA}). This is the only
approach that easily uncovers the galaxy distribution in the ZOA, 
allows the confirmation of implied connections and uncovers new
connections behind the Milky Way.

From the analysis of the Dwingeloo survey and the shallow Parkes MB
ZOA survey, it can be maintained that no Andromeda-like or other \HI-rich
galaxy is lurking undetected behind the deepest
extinction layers of the Milky Way (although gas-poor, early-type
galaxies might, of course, still remain hidden, like the recently
discovered very local dwarf behind the Galactic bulge,
Sect.~\ref{Sag}). The census of dynamically important, \HI-rich nearby
galaxies whose gravitational influence could significantly impact
peculiar motion of the Local Group or its internal dynamics is now
complete -- at least for objects whose signal is not drowned within
the strong Galactic \HI\ emission. Simulations are currently being
devised to investigate what kind of \HI\ galaxies -- whose signals
lie within the frequency range of the Milky Way's \HI\ -- could still 
have been missed.

\section{X-ray surveys} \label{Xray}
The X-ray band potentially is an excellent window for studies of
large-scale structure in the Zone of Avoidance, because the Milky Way
is transparent to the hard X-ray emission above a few keV, and because
rich clusters are strong X-ray emitters. Since the X-ray luminosity is
roughly proportional to the cluster mass as $L_X \propto M^{3/2}$ or
$M^2$, depending on the still uncertain scaling law between the X-ray
luminosity and temperature, massive clusters hidden by the Milky Way
should be easily detectable through their X-ray emission.

This method is particularly attractive, because clusters are primarily
composed of early-type galaxies which are not recovered by IRAS galaxy
surveys (Sect.~\ref{fir}) or by systematic \HI\ surveys
(Sect.~\ref{hi}). Even in the NIR, the identification of early-type
galaxies becomes difficult or impossible at the lowest Galactic
latitudes because of the increasing extinction and crowding problems
(Sect.~\ref{nir}).  Rich clusters, however, play an important role in
tracing large-scale structures because they generally are located at the
center of superclusters and Great Wall-like structures. They mark the
density peaks in the galaxy distribution and -- with the very high
mass-to-light ratios of clusters -- the deepest potential wells within these
structures.  Their location within these overdensities will help us
understand the observed velocity flow fields induced by these
overdensities.

The X-ray all-sky surveys carried out by Uhuru, Ariel V, HEAO-1 (in
the $2 - 10$~keV band) and ROSAT ($0.1 - 2.4$~keV) provide an optimal
tool to search for clusters of galaxies at low Galactic
latitude. However, confusion with Galactic sources such as X-ray
binaries and Cataclysmic Variables may cause serious problems,
especially in the earlier surveys (Uhuru, Ariel V and HEAO-1) which had
quite low angular resolution.  And although dust extinction and
stellar confusion are unimportant in the X-ray band, photoelectric
absorption by the Galactic hydrogen atoms -- the X-ray absorbing
equivalent hydrogen column density -- does limit detections close to
the Galactic Plane. The latter effect is particularly severe for the
softest X-ray emission, as \eg observed by ROSAT ($0.1 - 2.4$~keV)
compared to the earlier $2 - 10$~keV missions. On the other hand, the
better resolution of the ROSAT All Sky Survey (RASS) compared to the
HEAO-1 survey will reduce confusion problems with Galactic sources as
happened, for example, in the case of the cluster A3627 (see below).

Until recently, the possibility of searching for galaxy
clusters behind the Milky Way through their X-ray emission has not 
been pursued in a systematic way, even though a large number of 
X-ray bright clusters are located at low Galactic latitudes 
(Fabian 1994): for instance, four of the seven most X-ray luminous
clusters in the 2--10 keV range, the Perseus, Ophiuchus,
Triangulum Australis, and PKS\,0745$-$191  clusters ($L_{\rm X}>10^{45}$ 
erg s$^{-1}$) lie at latitudes below $|b|<20^{\circ}$ (Edge \etal 1990). 

A first attempt to identify galaxy clusters in the ZOA through their
X-ray emission had been made by Jahoda and Mushotzky in 1989. They
used the HEAO-1 all-sky data to search for X-ray-emission of a
concentration of clusters or one enormous cluster that might help
explain the shortly before discovered large-scale deviations from the
Hubble flow that were associated with the Great Attractor.
Unfortunately, this search missed the 6$^{th}$ brightest cluster A3627
in the ROSAT X-ray All Sky Survey (B\"ohringer \etal 1996, Tamura
\etal 1998) which had been identified as the most likely candidate for
the predicted but unidentified core of the Great Attractor
(Kraan-Korteweg \etal 1996).  A3627 was not seen in the HEAO-1 data
because of the low angular resolution and the confusion with the
neighbouring X-ray bright, Galactic X-ray binary 1H1556$-$605 (see
Fig.~8 and 9 in B\"ohringer \etal 1996).

\subsection{CIZA: Clusters in the Zone of Avoidance}

Since 1997, a group led by Ebeling (Ebeling \etal 1999, 2000) have
systematically searched for bright X-ray clusters of galaxies at $|b|
< 20\deg$. Starting from the ROSAT Bright Source Catalog (BSC, Voges
\etal 1999) which lists the 18811 X-ray brightest sources detected in
the RASS, they apply the following criteria to search for clusters:
(a) $|b|<20^{\circ}$, (b) an X-ray flux above $S > 5\times 10^{-12}$
erg cm$^{-2}$ s$^{-1}$ (the flux limit of completeness of the ROSAT
BCS), and (c) a spectral hardness ratio. Ebeling \etal demonstrated in
1998 that the X-ray hardness ratio is very effective in discriminating
against softer, non-cluster X-ray sources. With these criteria they
selected a candidate cluster sample which, although at this point still
highly contaminated by non-cluster sources, contains the final CIZA
cluster sample.

They first cross-identified their 520 cluster candidates against NED
and SIMBAD, and checked unknown ones on the Digitized Sky Survey.  The
new cluster candidates, including known Abell clusters without
photometric and spectroscopic data, were imaged in the $R$ band,
respectively in the $K'$ band in regions of high extinctions. With the
subsequent spectroscopy of galaxies around the X-ray position, the
real clusters could be confirmed.

Time and funding permitting, the CIZA team plans to extend their
cluster survey to lower X-ray fluxes ($2-3\times 10^{-12}$ erg 
cm$^{-2}$ s$^{-1}$), the aim being a total sample of 200 X-ray selected
clusters at $|b| < 20\deg$.

\begin{figure}[t]
\begin{center}
\hfil \psfig{file=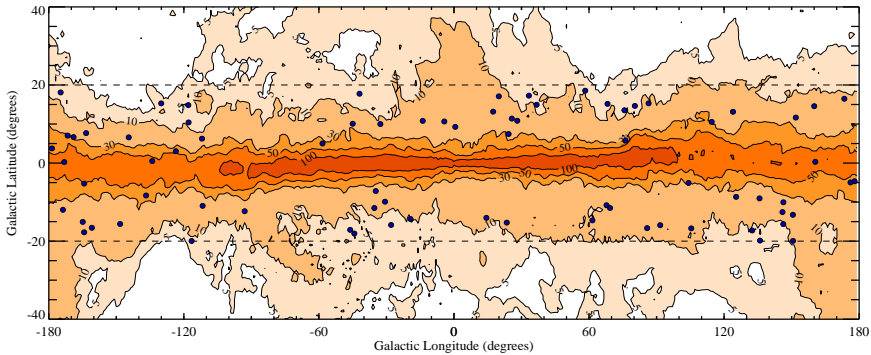,width=12cm} \hfil
\caption
{Distribution in Galactic coordinates of the 76 by Ebeling \etal
(2000) so far spectroscopically confirmed X-ray clusters (solid dots)
of which 80\% were previously unknown. Superimposed are Galactic HI
column densities in units of $10^{20}$ cm$^{-2}$ (Dickey \& Lockman
1990).  Note that the region of relatively high absorption ($N_{\rm HI}
> 5 \times 10^{21}$ cm$^{-2}$) actually is very narrow and that
clusters could be identified to very low latitides.
}
\label{xray}
\end{center}
\end{figure}

So far, 76 galaxy clusters were identified within $|b| < 20\deg$ of
which 80\% were not known before. Their distribution is displayed in
Fig.~\ref{xray} (reproduced from Ebeling \etal 2000). 14 of these
clusters are relatively nearby ($z \le 0.04$), and one was uncovered
within the Perseus-Pisces chain at a latitude of only $b = 0\fdg3$.

\subsection{Conclusions}
With the discovery of 76 clusters so far, of which only 20\% were known
before, Ebeling \etal (2000) have proven the strength of the method to
use X-ray criteria to search for galaxy clusters in the ZOA. As
mentioned in the introduction to this section, this approach is 
complementary to searches at other wavelengths which all fail to
uncover galaxy clusters at very low Galactic latitudes. 

Having used the ROSAT BSC to select their cluster candidates,
the CIZA collaboration wants to combine their final cluster sample with
other X-ray selected cluster samples from the RASS, such as the ROSAT
Brightest Cluster Sample at $|b|\ge 20^{\circ}$ and $\delta \ge 0\deg$
(Ebeling \etal 1998) and the REFLEX sample at $|b|\ge 20\deg$
and $\delta \le 2.5\deg$ (B\"ohringer \etal in prep.). The resulting,
all-sky cluster list will be ideally suited to study large-scale
structure and the connectivity of clusters across the Galactic Plane.

\section{Statistical reconstructions} \label{recon}
Where the Zone of Avoidance cannot be observed directly, the
alternative is to reconstruct the structure in a statistical way.
Corrections for unobserved regions in catalogs were done, somewhat
ad-hoc, by populating the ZOA uniformly according to the mean density,
or by interpolating the structure below and above the Galactic Plane
(\eg Lynden-Bell \etal 1989; Yahil \etal 1991; Strauss \etal 1992;
Hudson 1992).  Other authors utilized statistical methods such as
Wiener filtering (see below) to recover an all-sky density field
(Lahav \etal 1994; Hoffman 1994; Fisher \etal 1995; Zaroubi \etal
1995).  The recovery of a signal from noisy and incomplete data is a
classic problem of inversion, common in problems of image processing.
A straightforward inversion is often unstable, and a regularization
scheme of some sort is essential in order to interpolate where data
are missing or noisy.  In the Bayesian spirit, one can use raw data and
a prior model to produce an ``optimal reconstruction''.  Using the
above principle, one can derive the Wiener filter (the ratio of signal
to signal+noise), which also follows from requiring minimum variance
(\eg Rybicki \& Press 1992).  The formalism of Wiener filtering is
given in Appendix A.

\subsection{Mask inversion using Wiener filtering in spherical 
harmonic analysis}\label{sph}

It is convenient to expand the galaxy distribution in a nearly
whole-sky survey in spherical harmonics.  This was applied to 2-D
(i.e. projected on the sky) samples (\eg Peebles 1973; Scharf \etal
1992) and to redshift and peculiar velocity surveys (\eg Reg\"os \&
Szalay 1989; Scharf \& Lahav 1993; Lahav \etal 1994; Fisher \etal
1994; Nusser \& Davis 1994; Fisher \etal 1995; Heavens \& Taylor
1995).  In projection, the density field over $4 \pi$ is expanded as a
sum:
$$
{\cal S }(\theta, \phi) =
\sum_{l} \sum_{m=-l}^{m=+l} a_{lm}\; Y_{lm}(\theta, \phi),
$$ 
where the $Y_{lm}$'s are the orthonormal set of spherical harmonics
and $\theta$ and $\phi$ are the spherical polar angles.

The problem of reconstructing large-scale structure behind the ZOA can
be formulated as follows: what are the full-sky, noise-free harmonic
coefficients $a_{lm}$ given the observed harmonics, the mask
describing the unobserved region, and a prior model for the
power-spectrum of fluctuations?  In a projected catalog the
observed harmonics $c_{lm,obs}$ (with the masked regions filled in
uniformly according to the mean) are related to the underlying `true'
whole-sky harmonics $a_{lm}$ by (see Peebles 1980)
$$
c_{lm, obs} = \sum_{l'} \sum_{m'} W_{ll'}^{mm'} \;
[ a_{l'm'} \;  + \; \sigma_{a} ],
$$
where the monopole term ($l'=0$) is excluded.  The Poisson shot-noise
$\sigma_a $ is added to the number-weighted harmonics $a_{lm}$'s.  The
noise variance is estimated as $\langle \sigma_a^2 \rangle = N$ 
(the mean number of galaxies per steradian, independent of $l$) The
harmonic transform of the mask, $W_{ll'}^{mm'}$, introduces
`cross-talk' between the different harmonics.

It can be shown (Lahav \etal 1994; Zaroubi \etal 1995) that the
solution of this inversion problem is
$$ 
{\hat {\bf a}} = {\bf F} {\bf W}^{-1} \; {\bf c}_{obs}, 
$$
where the vectors ${\bf a}$ and ${\bf c}_{obs} $ represent the sets of
harmonics $\{ a_{lm} \} $ and $\{ c_{lm, obs} \}$, with the diagonal
Wiener matrix
$$
{\bf F} = diag \Big\{ { \langle a_l^2\rangle_{th} \over 
{ \langle a_l^2\rangle_{th} + \langle \sigma_a^2 \rangle }} \Big\}.
$$
Here $\langle a_l^2\rangle_{th}$ is the cosmic variance in the
harmonics, which depends on the power-spectrum.  In the special case
of an underlying Gaussian field the most probable field, the mean field,
and the minimum variance Wiener filter all are identical.  The scatter
in the reconstruction can be written analytically for Gaussian random
fields.

Even if the sky coverage is $4 \pi$ (${\bf W} = {\bf I}$), the Wiener
filter is essential to reveal the optimal underlying continuous
density field, cleaned of noise.  In the absence of other prior
information on the location of clusters and voids, the correction
factor is `isotropic' per $l$, i.e. independent of $m$. So, in the case
of full sky coverage, only the amplitudes are affected by the
correction, but not the relative phases.  For example, the dipole
direction is not affected by the shot-noise, only its amplitude.  But
if the sky coverage is incomplete, both the amplitudes and the phases
are corrected.  The reconstruction also depends on number of observed
and desired harmonics.  Note also that the Wiener method is {\it
non}-iterative.

\subsection{Reconstruction of the projected IRAS 1.2~Jy galaxy distribution}

Lahav \etal (1994) applied the method to the sample of IRAS galaxies
brighter than 1.2~Jy which includes 5313 galaxies, and covers 88\%
of the sky.  This incomplete sky coverage is mainly due to the Zone
of Avoidance, which they modelled as a `sharp mask' at Galactic
latitude $|b| < 5\deg$.  The mean number of galaxies is $N
\sim 400 $ per steradian, which sets the shot-noise, $\langle
\sigma_a^2 \rangle$.  As the model for the cosmic scatter $\langle
a_l^2 \rangle_{th}$, they adopted a fit to the observed power spectrum
of IRAS galaxies.

\begin{figure}[p]
\begin{center}
\hfil \psfig{file=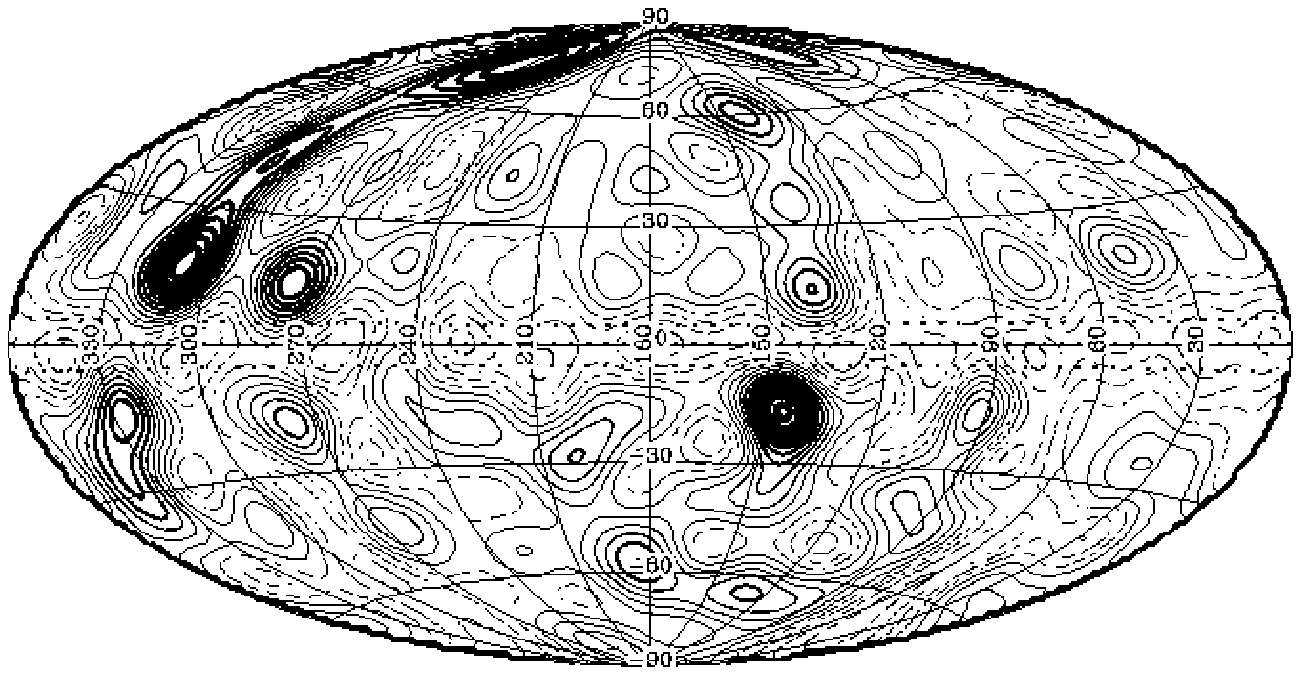,angle=0,width=12cm} \hfil
\caption
{Harmonic expansion ($ 1 \le l \le 15$) of the projected raw
IRAS 1.2 Jy data in Galactic Aitoff projection. Note that this
projection has the Anticenter (not the Galactic Bulge) at its
origin.  Regions not observed,
in particular $|b| < 5$ (marked by dashed lines), were left empty.
The contour levels of the projected surface number density are in
steps of 100 galaxies per steradian (the mean projected density is
$N \sim 400$ galaxies per steradian).  
(From Lahav \etal 1994.)}
\label{2DWienera}
\end{center}
\begin{center}
\hfil \psfig{file=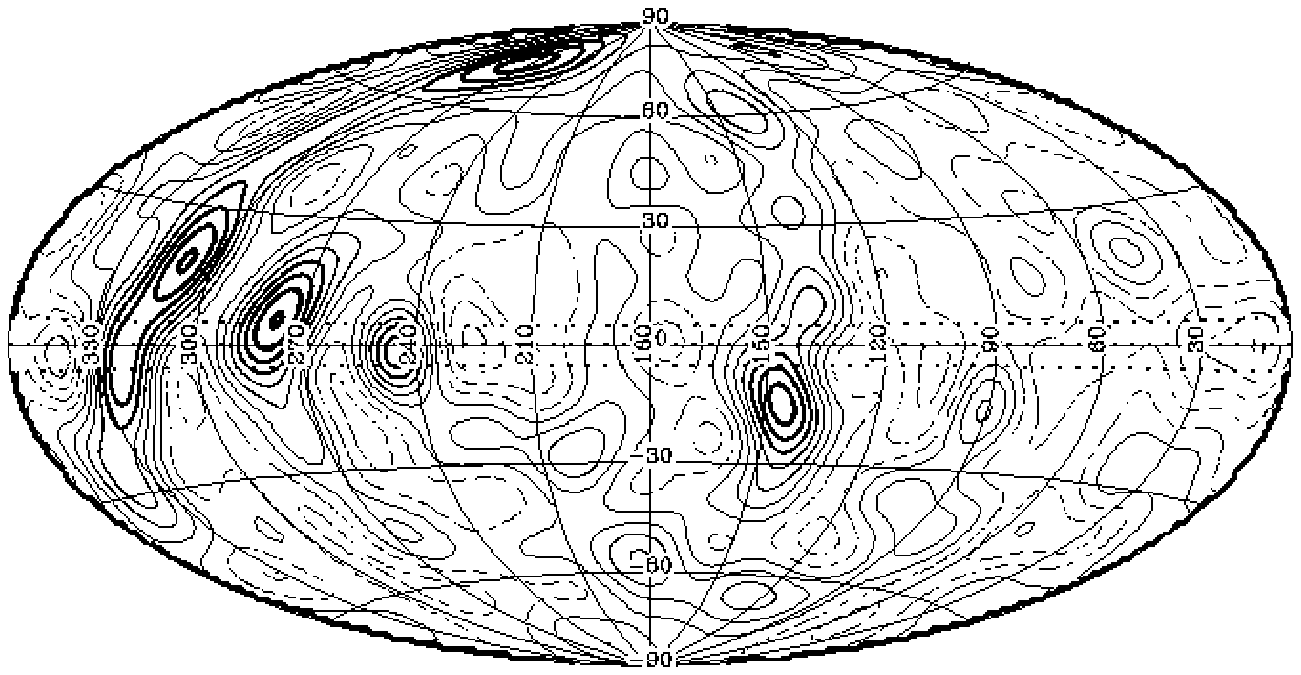,angle=0,width=12cm} \hfil
\caption
{
A whole-sky Wiener reconstruction of Fig.~15.  The
reconstruction corrects for incomplete sky coverage, as well as for
the shot-noise.  The reconstruction indicates that the Supergalactic
Plane is connected across the Galactic Plane at Galactic longitude
$\ell \sim 135\deg$ and $ \ell \sim 315\deg$. The Puppis cluster 
stands out at the Galactic Plane at $\ell \sim 245\deg$. The 
horizontal dashed lines at $b = \pm 5\deg$ mark the major 
ZOA in the IRAS sample. The contour levels are as in Fig.~15.  
(From Lahav \etal 1994.)}
\label{2DWienerb}
\end{center}
\end{figure}

Fig.~\ref{2DWienera} shows the reconstruction of the projected IRAS
1.2~Jy sample.  The ZOA was left empty, and it clearly `breaks' the
possible chain of the Supergalactic Plane and other
structures. Fig.~\ref{2DWienerb} shows our optimal reconstruction for
$ 1 \le l \le 15$.  Now the structure is seen to be connected across
the ZOA, in particular in the regions of Centaurus/Great Attractor
($\ell \sim 315\deg$), Hydra ($\ell \sim 275\deg$) and Perseus-Pisces
($\ell \sim 315\deg$), confirming the connectivity of the
Supergalactic Plane.  Note that Figs.~\ref{2DWienera} and
\ref{2DWienerb} are shifted by $180\deg$ compared to the earlier sky
projections. We also see the Puppis cluster ($\ell \sim 245\deg$)
recovered behind the Galactic Plane. This cluster was noticed in
earlier harmonic expansion (Scharf \etal 1992) and further studies
(Lahav \etal 1993).  The other important feature of this
reconstruction is the suppression of shot noise all over the sky. This
is particularly important for judging the reality of clusters and
voids.

Comparison of this Wiener reconstruction with the one applied (using a
$4 \pi$ Wiener filter) to the IRAS sample, in which the ZOA was filled
in `by hand' across the Galactic Plane (Yahil \etal 1991), shows good
agreement.  By testing the method on $N$-body simulations (where the
whole `sky' true harmonics are known) it was found, that for masks
larger than $|b|=15\deg$, it is difficult to recover the unobserved
structure.  In this case extra-regularization was required.  The
success of the method depends on the interplay of three angular
scales: the width of the mask, the desired resolution ($\pi/l_{max}$)
and the physical correlation of structure.  It is important to note
that, as opposed to ad-hoc smoothing schemes, the smoothing due to a
Wiener filter is determined by the sparseness of data relative to the
expected signal.

\subsection{3-D  reconstruction}

Fisher \etal (1995) generalized the 2-D Wiener reconstruction to 3-D
(i.e. for redshift catalogs) by expanding the density field $\rho$ in
redshift-space in terms of spherical harmonics, $Y_{lm}$, and radial
Bessel functions $j_l$:
$$
\rho(\bfs) = 
\sum\limits_{l=0}^{l_{\rm max}}
\sum\limits_{m=-l}^{+l}
\sum\limits_{n=1}^{n_{\rm max}(l)}\, 
C_{ln}\, \rho_{lmn}\, j_l(k_n s)\, Y_{lm}(\hat\bfs).
\qquad 
$$
The discrete $k_n$'s are chosen according to the boundary conditions,
so as to make the set orthogonal. This process is analogous to Fourier
decomposition, but uses instead a set of spherical basis
functions. The data from a redshift catalog can be seen as a set of
$N$ discrete points, $\bfs_i$, each giving the direction and redshift
of a galaxy. These are used to estimate the underlying density field
in redshift space, $\hat\rho^\bfS(\bfs)$, expanded as in the above
equation. Here, $C_{ln}$ are normalization constants, while the
harmonic coefficients are given by
$$
\hat\rho_{lmn}^{\bfS} = 
\sum\limits_{i=1}^N {{1}\over{\phi(s_i)}}
j_l(k_ns_i) Y^\ast_{lm}(\bf {\hat\bfs_i} ), \qquad
$$
where $\phi(s_i)$ is the spherical selection function of the survey,
evaluated at the radius of the $i^{th}$ galaxy.  The real-space
density, velocity and potential fields are then reconstructed using
linear theory and a Wiener filter.

In analogy with the 2-D case, a mask (\eg to describe the ZOA) can be
formulated in the 3-D harmonic formalism (Heavens \& Taylor 1995;
Schmoldt \etal 2000).  However, for the simplicity of the analysis,
Fisher \etal (1995) first populated the ZOA with mock galaxies
according to the procedure of Yahil \etal (1991), and then applied
the reconstruction over the full sky.  Webster \etal (1997)
and Schmoldt \etal (2000) extended this analysis and presented
detailed maps of the reconstructed fields, as well as optimal
determinations of the Local Group dipole and bulk flows.  Lahav \etal
(2000) used Wiener filtering to study the the extent of the
Supergalactic Plane.  Bistolas (1998) has done a similar Wiener
reconstruction in Cartesian coordinates (see discussion below).
Saunders \etal (2000a) have extended the spherical harmonics and Wiener
approach to represent non-linear clustering by describing the density
field as drawn from a log-normal probability distribution function.
They recently applied it to the PSCz IRAS catalog (taking into account
the detailed IRAS mask, see Fig.~\ref{BTP}). This type of analysis,
when applied to other surveys at low latitude, can potentially provide
the most detailed and self-consistent map of the ZOA.

\subsection{POTENT reconstruction of the ZOA}\label{potent}

The POTENT method (Bertschinger \& Dekel 1989; Dekel 1994) recovers
the smoothed fluctuations field of potential, velocity and mass
density from observed radial peculiar velocities of galaxies. The
velocity field is recovered under the assumption of potential flow,
${\bf v ({\bf r}) } = - \nabla \Phi ({\bf r}) $.  The potential can
thus be calculated by integrating the radial velocity along radial
rays. Differentiating $\Phi$ in the transverse directions recovers the
two missing velocity components.  The underlying mass-density
fluctuation $\delta$ is then derived in linear theory from $\delta
({\bf r}) = - \Omega^{-0.6} \nabla \cdot {\bf v}({\bf r})$ (or from
non-linear extensions).  This method is very useful for exploring the
ZOA as the peculiar velocity field responds to the entire mass
distribution, regardless of the `unseen' distribution of light.
However, due to the heavy smoothing required by this method, only
structures on large scales (\eg superclusters) can be mapped.
Individual (massive) nearby galaxies that may perturb the dynamics in
the vicinity of the Local Group cannot be uncovered in this manner.
Kolatt \etal (1995) used the POTENT method to specifically predict
the mass distribution behind the ZOA (see Fig.~\ref{kolatt}).  Some of
their predictions are summarized below.

Zaroubi \etal (1999) analyzed the peculiar velocity field using Wiener
filtering.  The reconstructed structures are consistent with those
extracted by the POTENT method. A comparison with the structures in
the distribution of IRAS 1.2~Jy galaxies yields a general agreement.

\begin{figure}[p]
\begin{center}
\hfil \psfig{file=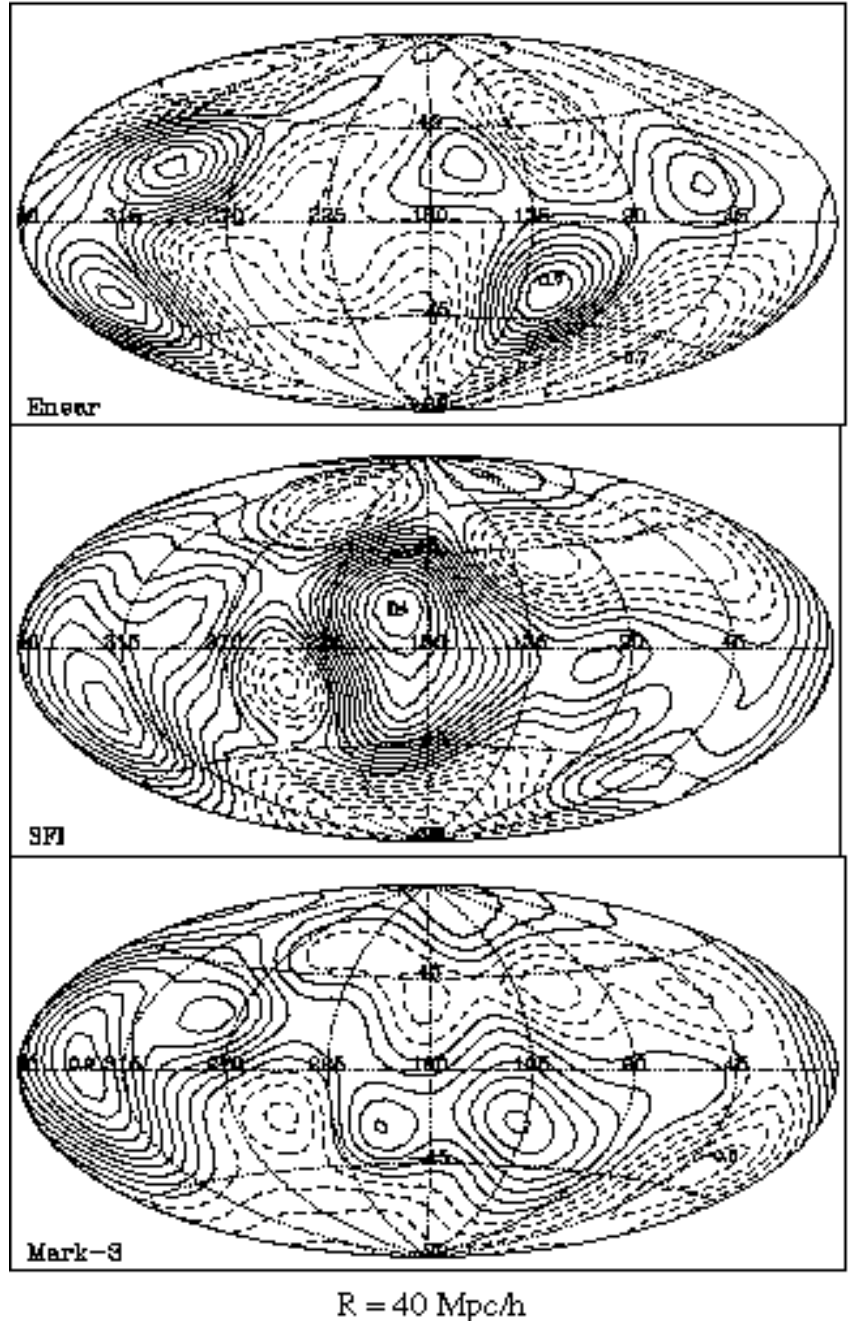,width=11.2cm} \hfil
\caption
{Aitoff projection in Galactic coordinates (centered
on the Galactic Anticenter) 
of reconstruction at $R=40 h^{-1}_100$ Mpc from three
velocity catalogs: Mark III, SFI and ENEAR.
The applied smoothing is with a Gaussian of $9~h^{-1}_100$ Mpc.
The Mark III and SFI reconstructions
are from Zaroubi, Hoffman \& Dekel (1999)  
for ENEAR from  Zaroubi (2000) and Zaroubi \etal (in preparation).
}
\label{zaroubi}
\end{center}
\end{figure}

Figure~\ref{zaroubi} shows an Aitoff projection in Galactic
coordinates of reconstrucion at $R=40~h^{-1}_100$ Mpc from the three
velocity catalogs: Mark III (Willick \etal 1997), 
SFI (Survey of Field Spirals in the I-band; da Costa
\etal 1996; Giovanelli \etal 1998) and ENEAR (the ESO Nearby Early-type 
Galaxies Survey; da Costa \etal 2000a, 2000b).  For further discussion of
the Wiener reconstruction of the ZOA from velocities see Hoffman
(2000).


\subsection{Predicted structures behind the ZOA}
Early reconstructions on relatively sparse data galaxy catalogs have
been performed within volumes out to $v \le$ 5000~\kms. Despite
heavy smoothing, they have been quite successful in 
pinpointing a number of important features. 

Scharf \etal (1992) applied spherical harmonics to the
2-dimensional IRAS PSC and noted a prominent cluster behind the ZOA in
Puppis ($\ell \sim 245\deg$) which was simultanously discovered as a
nearby cluster through \HI-observations of obscured galaxies in that
region by Kraan-Korteweg \& Huchtmeier (1992). It was analyzed 
further by Lahav \etal (1993).

Hoffman (1994) predicted the Vela supercluster at ($280\deg,
6\deg, 6000$~\kms), using 3-dimensional Wiener filter reconstructions
on the IRAS 1.9~Jy redshift catalog (Strauss \etal 1992). It was
discovered observationally just a bit earlier by Kraan-Korteweg \&
Woudt (1993).

\begin{figure}[ht]
\begin{center}
\hfil \psfig{file=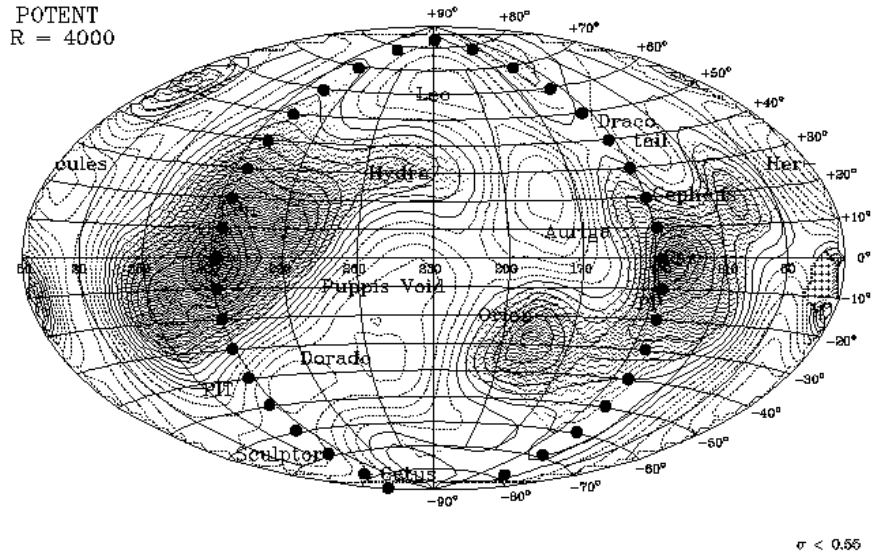,angle=-90,width=12cm} \hfil
\caption
{The mass-density fluctuation field in a shell at 4000~\kms\
as determined with POTENT from peculiar velocity data. The 
density is smoothed by a three-dimensional Gaussian of radius 
1200~\kms. Contour spacings are $\Delta\delta = 0.1$, with 
$\delta=0$ the heavy contour. Compared to Fig.~1 and 3,
this Aitoff projection is displaced by $\Delta \ell =
50\deg$. The Supergalactic Plane is indicated (solid dots).
(Figure 1b from Kolatt \etal 1995)}
\label{kolatt}
\end{center}
\end{figure}

Using POTENT analysis, Kolatt \etal (1995) predicted the
center of the Great Attractor overdensity -- its density peak -- to
lie behind the ZOA at ($320\deg, 0\deg, 4500$~\kms; see 
Fig.~\ref{kolatt}). Shortly thereafter, Kraan-Korteweg \etal (1996)
unveiled the cluster A3627 as being very rich and massive and at the
correct distance. It hence is the most likely candidate for the
central density peak of the GA.

POTENT reconstructions have been applied to denser galaxy samples
covering larger volumes (8-10000~\kms) with smoothing scales of the
order of 500~\kms\ (compared to 1200~\kms). It therefore seemed of
interest to see whether these reconstructions find evidence for
unknown major galaxy structures in the ZOA at higher redshifts.

The currently most densely-sampled, well-defined galaxy redshift
catalog is the Optical Redshift Survey (Santiago \etal 1995).
However, this catalog is limited to $|b| \ge 20\deg$ and the
reconstructions (see Baker \etal 1998) within the ZOA are strongly
influenced by 1.2~Jy IRAS Redshift Survey data and a mock galaxy
distribution in the inner ZOA.  We will therefore concentrate on
reconstructions based on the 1.2~Jy IRAS Redshift Survey only. 

In the following, the structures identified in the ZOA by (a) Webster
\etal (1997) using Wiener filter plus spherical harmonics and linear
theory and (b) Bistolas (1998), who applied a Wiener filter plus linear
theory and constrained realizations, will be discussed and compared to
observational data.  Figure~2 in Webster \etal displays the
reconstructed density fields on shells of 2000, 4000, 6000 and
8000~\kms; Fig.~5.2 in Bistolas displays the density fields in the ZOA
from 1500 to 8000~\kms\ in steps of 500~\kms.

The reconstructions by Webster \etal 1997 clearly show the recently
identified nearby cluster at ($33\deg,5\deg-15\deg$,1500\kms; see
Sect.~\ref{optred}), whereas Bistolas reveals no clustering in the
region of the Local Void out to 4000~\kms. At the same longitudes, the
clustering at 7500~\kms\ is seen by Bistolas, but not by Webster \etal
The Perseus-Pisces chain is strong in both reconstructions, and the
second Perseus-Pisces arm -- which folds back at $\ell\sim 95\deg$ -- is
clearly confirmed.  Both reconstructions find the Perseus-Pisces
complex to be very extended in space, \ie\ from 3500~\kms\ out to
9000~\kms.  Whereas the GA region is more prominent compared to
Perseus-Pisces in the Webster \etal reconstructions, the signal of the
Perseus-Pisces complex is considerably stronger than the GA in
Bistolas, where it does not even reveal a well-defined central density
peak. Both reconstructions find no evidence for the suspected
cluster around PKS1343$-$601, but its signal could be hidden in the
central (A3627) density peak due to the smoothing.  While the
Cygnus-Lyra complex ($60\deg-90\deg, 0\deg, 4000$~\kms) discovered by
Takata \etal (1996) stands out clearly in Bistolas, it is not evident
in Webster \etal  Both reconstructions find a strong signal for the
Vela SCL ($285\deg, 6\deg, 6000$\kms), labelled as HYD in Webster \etal
1997. The Cen-Crux cluster identified by Woudt (1998) is evident in
Bistolas though less distinct in Webster \etal A suspected connection
at ($\ell,v) \sim (345\deg, 6000$\kms; see middle panel of
Fig.~\ref{vslice}) is supported by both methods.  The Ophiuchus cluster
just becomes visible in the most distant reconstruction shells
(8000~\kms).

\subsection{Conclusions}
Not all reconstructions find the same features, and when they do, the
prominence of the density peaks as well as their locations in space do
vary considerably. At velocities of $\sim 4000$~\kms\ most of the
dominant structures happen to lie close to or within the ZOA while at
larger distances, clusters and voids seem to be more homogeneously
distributed over the whole sky. Out to 8000~\kms\ none of the
reconstructions predict any major structures which are not mapped or
suggested from observational data. Thus, no major surprises seem to
remain hidden in the ZOA. The various multi-wavelength explorations of
the Milky Way will soon be able to verify this. Still, the combination
of both the reconstructed potential fields and the observationally
mapped galaxy distribution will lead to estimates of the cosmological
parameters $\Omega_0$ and $b$.

\section{Discussion}

In the last decade, enormous progress has been made in unveiling the
extragalactic sky behind the Milky Way. At optical wavebands, the
entire ZOA has been systematically surveyed. It has been shown that
these surveys are complete for galaxies larger than $D^o = 1\farcm3$
(corrected for absorption) down to extinction levels of ${A_B} =
3\fm0$. Combining these data with previous ``whole-sky'' maps reduces
the ``optical ZOA'' by a factor of about 2 - 2.5, which allows an
improved understanding of the velocity flow fields and the total
gravitational attraction on the Local Group. Various previously
unknown structures in the nearby Universe could be mapped in this way.

At higher extinction levels, other windows to the ZOA become more
efficient in tracing the large-scale structures.  Very promising in
this respect are the current near-infrared surveys which find galaxies
down to latitudes of $|b| \sim 1\fdg5$. Source confusion will remain a
problem at low Galactic latitude which may be overcome by introducing
novel statistical methods such as Artificial Neural Networks. The
systematic \HI\ surveys detect gas-rich spiral galaxies all the way
across the Galactic Plane -- slightly hampered only at very low
latitudes ($|b|\la 1\fdg0$) because of the numerous continuum sources.
These studies have already shown that no unknown dynamically
important, \HI-rich nearby galaxies whose gravitational influence
could significantly impact the internal dynamics and the peculiar
motion of the Local Group are hidden by the Milky Way. In addition,
the deep ZOA \HI\ surveys can be merged with the lower-sensitivity
whole-sky \HI-surveys currently in progress. The ``Behind the Plane''
survey resulted in a reduction from 16\% to 7\% of the ``FIR ZOA''
which soon should provide improved values of the dipole direction and
convergence from IRAS data. In addition, new indications of possible
hidden massive clusters behind the Miky Way are now forthcoming from
the CIZA project -- although again an ``X-ray ZOA'' will remain due to
the absorption of X-ray radiation by the thick gas layer close to
the Galactic Plane.

A difficult task is still awaiting us, \ie to obtain a detailed
understanding of the selection effects inherent to the various
methods. Quantifying the selection effects is crucial for any optimal
reconstruction method (\eg Wiener) which attempts to merge the
different data sets in an unbiased way. This is extremely important if
we want to use these data for quantitative cosmography. Moreover, we
need a better understanding of the effects of obscuration on the
observed properties of galaxies identified through the dust layer (at
all wavelengths), in addition to an accurate high-resolution,
well-calibrated map of the Galactic extinction.

Despite the fact that our knowledge about the above issues is as yet
limited, a lot can and has been learned from ZOA research. This is
evident, for instance, from the detailed and varied investigations of
the Great Attractor region. Mapping the GA and understanding the
massive overdensity inferred from peculiar velocity fields had
remained an enigma due the fact that the major and central part of
this extended density enhancement was largely hidden by the obscuring
veil of the Milky Way. The results from the various ZOA surveys now
clearly imply that the Great Attractor is, in fact, a nearby
``Great-Wall'' like supercluster, starting at the nearby Pavo cluster
below the GP, moving across the massive galaxy cluster A3627 toward
the shallow overdensity in Vela at 6000~\kms. The cluster A3627 is the
dominant central component of this structure, similar to the Coma
cluster in the (northern) Great Wall. Whether a second massive cluster
around PKS1343$-$601 is part of the core of the GA remains
uncertain.

\small
\acknowledgements
The enthusiastic collaborations of our colleagues in the exploration of the
galaxy distribution behind the Milky Way is greatly appreciated. These
are P.A. Woudt, C. Salem and A.P. Fairall with deep optical searches,
C. Balkowski, V. Cayatte, A.P. Fairall, P.A. Henning with redshift
follow-ups of optically identified galaxies, A. Schr\"oder and
G.A. Mamon in the exploration of DENIS images at low Galactic
latitude, W.B. Burton, P.A. Henning, and A. Rivers in the
northern ZOA HI-survey (DOGS) and the HIPASS ZOA team members
L. Staveley-Smith , R.D. Ekers, A.J. Green, R.F. Haynes, P.A. Henning,
S. Juraszek, M. J. Kesteven, B. Koribalski, R.M. Price, E. Sadler and
A. Schr\"oder in the southern ZOA survey.

Particular thanks go to P.A. Woudt for his valuable suggestions, to
W. Saunders for preparing Fig.~\ref{BTP}, to A. Schr\"oder and
G. Mamon for their comments on the NIR section, and to H. Ebeling for
his input with regard to the X-ray section and Fig.~\ref{xray}.

This research has made use of the NASA/IPAC Extragalactic Database (NED)
which is operated by the Jet Propulsion Laboratory, Caltech, under
contract with the National Aeronautics and Space Administration,
as well as the Lyon-Meudon Extragalactic Database (LEDA),
supplied by the LEDA team at the Centre de Recherche Astronomique de
Lyon, Observatoire de Lyon.

\newpage

\noindent {\bf Appendix A: Wiener Filtering } \\

\noindent
Here we give a brief review of the Wiener (1949) filter (WF) technique;
the reader is referred to Lahav etal. (1994), Zaroubi et al. (1995)
and Rybicki \& Press (1992) for further details. Let us assume that we
have a set of measurements, $\{d_\alpha\}\ (\alpha=1,2,\dots N)$ which
are a linear convolution of the true underlying signal, $s_\alpha$,
plus a contribution from statistical noise, $\epsilon_\beta$, such
that
$$
d_\alpha = {\cal R}_{\alpha\beta}\left [ s_\beta +
\epsilon_\beta\right],
\qquad 
\eqno(A1) 
$$
where ${\cal R}_{\alpha\beta}$ is the response or ``point spread''
function (summation convention assumed).  Notice that we have assumed
that the statistical noise is present in the underlying field and
therefore is convolved by the response function.

The WF is the {\it linear} combination of the observed data which is
closest to the true signal in a minimum variance sense. More
explicitly, the WF estimate is given by $s_\alpha (WF) =
F_{\alpha\beta}\, d_\beta$ where the filter is chosen to minimize
\mbox{$\langle |s_\alpha(WF)-s_\alpha|^2\rangle$}. It is straightforward to
show that the WF is given by 
$$
F_{\alpha\beta} = \langle s_\alpha d_\gamma \rangle
\langle d_\gamma d_\beta^\dagger \rangle^{-1},\qquad 
\eqno(A2) 
$$
where
$$
\langle s_\alpha d_\beta^\dagger\rangle = {\cal R}_{\beta\gamma}\,
\langle s_\alpha s^\dagger_\gamma\rangle,\qquad 
\eqno(A3) 
$$
$$
\langle d_\alpha d_\beta^\dagger\rangle = 
{\cal R}_{\alpha\gamma}\, {\cal R}_{\beta\delta}\, \left[ 
\langle s_\gamma s^\dagger_\delta\rangle
+ \langle \epsilon_\gamma\epsilon^\dagger_\delta\rangle\right]. \qquad 
\eqno(A4) 
$$
In the above equations, we have assumed that the signal and noise are
uncorrelated. From equation A4, it is clear that, in order to implement
the WF, one must construct a {\it prior} which depends on the variance
of the signal and noise.

The dependence of the WF on the prior can be made clear by defining
signal and noise matrices given by $S_{\alpha\beta}=\langle s_\alpha
s^\dagger_\beta\rangle$ and $N_{\alpha\beta}=\langle \epsilon_\alpha
\epsilon^\dagger_\beta\rangle$. With this notation, we can rewrite
equation A4 as
$$
{\bf s}(WF) = {\bf S} \left[ {\bf S} + {\bf N}\right]^{-1} {\bf {\cal
R}}^{-1} {\bf d}. \qquad 
\eqno(A5) 
$$
Formulated in this way, we see that the purpose of the WF is to
attenuate the contribution of low signal-to-noise ratio data and
therefore regularize the inversion of the response function. The
derivation of the WF given above follows from the sole requirement of
minimum variance and requires only a model for the variance of the
signal and noise. The WF can also be derived using the laws of
conditional probability if the underlying distribution functions for
the signal and noise are assumed to be Gaussian; in this more
restrictive case, the WF estimate is, in addition to being the minimum
variance estimate, also both the maximum a {\it posterior} estimate
and the mean field. For Gaussian fields, the mean WF field can be
supplemented with a realization of the expected scatter about the mean
field to create a realization of the field; this is the heart of the
``constrained realization'' approach described in Hoffman \& Ribak
(1991).  A generalization to non-Gaussian fields is given in Sheth
(1995).
\end{document}